# The concept 'indistinguishable'[1]

# Simon Saunders

**Abstract.** The concept of indistinguishable particles in quantum theory is fundamental to questions of ontology. All ordinary matter is made of electrons, protons, neutrons, and photons and they are all indistinguishable particles. Yet the concept itself has proved elusive, in part because of the interpretational difficulties that afflict quantum theory quite generally, and in part because the concept was so central to the discovery of the quantum itself, by Planck in 1900; it came encumbered with revolution.
I offer a deflationary reading of the concept 'indistinguishable' that is identical to Gibbs' concept 'generic phase', save that it is defined for state spaces with only *finitely*-many states of bounded volume and energy (finitely-many orthogonal states, in quantum mechanics). That, and that alone, makes for the difference between the quantum and Gibbs concepts of indistinguishability.
This claim is heretical on several counts, but here we consider only the content of the claim itself, and its bearing on the early history of quantum theory rather than in relation to contemporary debates about particle indistinguishability and permutation symmetry. It powerfully illuminates that history.

## 1. Introduction

If there is any consensus as to what particle indistinguishability means, it is its formal expression in quantum mechanics: particles must have exactly the same mass, spin, and charge, and their states must be *symmetrised*, yielding either symmetric or antisymmetric wave-functions. This much was set in stone by Dirac almost a century ago. But the implications of this depends on what the wave-function means, and what entanglement means (because symmetrisation of the state usually involves entanglement), and on what a particle is and what kinds of representations of the permutation group are acceptable and how they should be motivated – and thereby, on a whole raft of interpretive questions in quantum mechanics and in philosophy of symmetry. To first seek common ground on all these topics would be a challenging task.[2]

Here we take an alternative approach: to consider the concept 'indistinguishable' *prior* to quantum mechanics proper; to consider the concept or concepts (by whatever name) that were actually in play leading up to Dirac's definitive formulation of the concept in 1926. Fairly directly, and uncontrovertibly, that takes us to a series of papers by Einstein on the quantum theory of the ideal monatomic gas, published in 1924-25 (which Dirac certainly studied), almost his last papers on quantum theory. It was for this gas theory, following Bose's method used in a new derivation of the Planck black-body radiation formula in 1924, that Bose-Einstein statistics was named.

In Einstein's second paper on the quantum theory of gases we find a structural correspondence between the old statistics and the new that reduces, in the dilute limit, to that between Boltzmann's statistics and the statistics defined using Gibbs' concept 'generic phase'– a concept introduced in the ultimate chapter of Gibbs' *Elementary Principles in*

---

[1] To appear in *Studies in History and Philosophy of Modern Physics*
https://doi.org/10.1016/j.shpsb.2020.02.003

[2] See Saunders (2003, 2013) for an attempt along these lines.

*Statistical Mechanics* in 1902. According to the generic phase, many-particle states differing only by particle interchange are numerically the same. It was designed to solve the Gibbs paradox and the closely connected puzzle of the failure of extensivity of the entropy function, and applied to what Gibbs called 'indistinguishable' particles, particles with the same state-independent properties. Indeed, the claim defended here is that the *only* distinction between the Gibbs and Bose-Einstein notions lies in the discretisation of the state-space by Einstein, following Bose, and the accompanying replacement of a volume measure (Lebesgue measure) on state space by a numerical count of discrete states. These were defined by Planck's constant: the number of 'elementary cells' of volume $h^3$ in the one-particle phase space. In the dilute limit, in which cells are sparsely occupied, the result is Gibbs' generic phase.[3] The identification of states related to one another by particle permutations is all that is needed to implement the indistinguishability concept, in the quantum case as in the classical.

This structural correspondence can hardly be gainsaid. However, it raises a certain difficulty, for neither Einstein nor anybody else at the time recognised it explicitly. Neither Einstein nor Dirac mentioned Gibb' concept at all, despite the evident connection with the extensivity puzzle (that Einstein did discuss) and the Gibbs paradox. The following year, in 1925, Schrödinger considered a number of definitions of the entropy that included the Bose-Einstein entropy, and in light of the extensivity puzzle; but only to reject the new gas theory, and again, he made no mention of Gibbs or the generic phase. The first to do so in connection with quantum statistics appears to have been Paul Epstein in his entry to Volume 2 of the *Commentary on the Scientific Writings of J. Willard Gibbs*, published in 1936 by Gibbs' alma mater, the University of Yale. But coming so late, it has rarely been consulted by historians of the earlier period, and it suffered from the presupposition common to almost every account of particle indistinguishability after Dirac's, that the quantum concept cannot possibly apply to classical statistical mechanics because *classical particles can always be distinguished by their trajectories*.[4] Yet as we shall see, it was exactly by endowing the light quantum with a phase space of its own, whereupon its states may change over time and so possess a trajectory, that Bose made his breakthrough in 1924.

There have been many studies of the history of the Planck radiation formula. However, apart from Thomas Kuhn's pioneering work, *Black-Body Theory and the Quantum Discontinuity 1894-1912*, most have considered it in the context of the broader history of quantum theory, from which vantage point it makes only an episodic appearance; and Kuhn, in a history terminating in 1912, did not consider the concept of indistinguishability at all. Those that have focused on the concept of indistinguishable particles without exception embrace the conventional wisdom that the quantum concept requires the absence of trajectories, or anything else that may keep track of individual particles; whereupon it follows that classical particles cannot possibly be indistinguishable. No wonder that in consequence they were unable to identify the quantum concept with Gibbs'.[5] Abraham Pais, in *Subtle is the Lord, the*

---

[3] An observation that appears to have first been made by Fujita (1991).
[4] Epstein (1936a p.557). Epstein also maintained that a continuous classical evolution, for a discrete state-space, requires Gibbs' specific phase (1936a p.564, 527, 1936b pp.485-91).
[5] Pesic (1991a) came close, but in a second publication affirmed that particles with trajectories cannot be indistinguishable in the quantum sense (Pesic 1991b p.975). Kastler (1983 p.617) mentioned the connection with Gibbs' generic phase, but only in passing, and insisted that particle indistinguishability is 'intimately related to their wave nature' (ibid, p.622). Monaldi (2009) argued that Gibbs' concept of generic phase was one of *two* roots to the indistinguishability concept in quantum theory, the other being statistical dependence; she does not remark on the Gibbs paradox and the lack of statistical independence already implied by it. According to Darrigol (1991 p.239), Gibbs was the first to introduce the indistinguishability concept, but only in the context of what he calls the 'molar' or 'holistic' approach to probability (essentially Gibbs' ensemble interpretation of probability, where each member of the ensemble of gases is defined by its *specific* phase); he

*Science and the Life of Albert Einstein*, is exemplary: according to Pais, Planck's early derivations of the black-body formula 'prefigured the Bose-Einstein counting' but 'cannot be justified by any stretch of the classical imagination'. In contrast, the distinguishability of particles by their initial conditions and the continuity of their motion is 'Boltzmann's first axiom of classical mechanics'.[6] Alexander Bach's more recent monograph *Classical Indistinguishable Particles* begins with the statement that the only intelligible classical concept of indistinguishability concerns the invariance of probability distributions that are blind to the motions of individual particles; for 'suppose they have trajectories, then the particles can be identified by them and are, therefore, not indistinguishable', a statement he proceeded to formalise mathematically and evidently took to be definitively proven.[7]

There is a related and widely-held view in the literature which connects even more directly with the ordinary meaning of the term: it is not only that indistinguishable quantum particles cannot have trajectories, it is that they cannot have states of their own at all – or if they do, those states must be the same. It was a short step to conclude that indistinguishable particles could not have precise positions and momenta.[8] There was moreover a clear link between this and the idea of energy elements or energy-units in the writings of Boltzmann, Planck, and Natanson. These entities had no positions or momenta by which they could differ, let alone trajectories or variations in energy; it seemed the only change they could undergo was emission and absorption (or 'containment') by other entities. Natanson has often been accredited, by historians, as the first to articulate the indistinguishability concept,[9] and on this point he was quite explicit: for something to count as indistinguishable in his sense, it must be impossible to 'lay hold' of it, or to 'trace it through its course'. It was not the Gibbs' concept, and nor was it the quantum concept.

Limitations in identifying entities at one time lead to inaccessibility of identity over time, the two go together. But these concepts of indistinguishability are all wide of the mark. The blue photon is perceptually distinguishable from the red photon and the two are heading in different directions in space, and may change in direction and frequency over time, and hence in momentum; photons may well have approximate trajectories. Yet their states are symmetrised, and they satisfy Bose-Einstein statistics. Their classical particle limit is a gas defined by Gibbs' generic phase: states related by particle interchange are identically the same. Indeed, it is they, the many-particle states related by permutations, that are indistinguishable in the epistemic sense, rather than the particles. Gibbs first used the word 'undistinguishable' with reference to many-particle states related by permutations; Dirac used 'indistinguishable' in the same way; but Dirac also spoke of permutations ('transitions') as indistinguishable, rather than states. In any case, following Dirac's work, some term was needed in quantum theory for particles whose states must be symmetrised; 'indistinguishable' is now common, 'identical particles' usually being reserved for particles that have the same

---

elsewhere insisted that the quantum concept of indistinguishability requires the absence of trajectories (1986 p.205-9).
[6] Pais (1982 p.371, p.63); the reference is to Boltzmann (1897 p.9). Pais shared with Bohr and his followers the belief that traditional ideas of objects and identity must be abandoned in quantum theory. For a detailed history of the identity concept in physics, sympathetic to this view, see French and Krause (2006) (which mentions Gibbs' concept of generic phase only once, and in passing, p.83).
[7] Bach (1997 p.8). Bach actually showed that microstates in the sense of specific phases cannot be invariant under permutations (if particles are impenetrable) – see Saunders (2006). His probability distributions are all defined on specific phase space (Gibbs did the same, but in the name of convenience, as we shall see).
[8] And must satisfy the Heisenberg's uncertainty relations, as claimed by e.g. Landau and Lifshitz (1958 p.204), Delbruck (1980 p.470), Pais (1982), and Kastler (1983 p.609).
[9] In a tradition beginning with Jammer (1966). Exceptions are Mehra and Rechenberg (1982 p.151), who mention him only in a footnote, and Pais, who does not mention him at all.

state-independent properties. Rarely can terminology in physics have been the source of so much misdirection.

Clarity on this point is not only an addition to the history of quantum theory, it throws a radical new light on it. Progress on black-body radiation more or less stalled, circa 1911, and awaited Bose's bold proposal of 1924 that the light quantum should have a state space of positions and momenta, just as does a material particle. That was entirely consistent with Gibbs' concept of the generic phase but completely at odds with the notion of light quanta as indistinguishable entities in the sense of Boltzmann, Planck, and Natanson. No wonder Bose and Einstein were slow to remark on the connection. What was immediately apparent to Einstein was that Bose's method could be applied to non-relativistic particles with mass, objects even more appropriately considered to have trajectories, resulting in a new equation of state and a new phase of matter (the Bose-Einstein condensate). If he realised that such particles, treated using Bose's method, could not possibly be statistically independent of each other, he gave no sign. It was only in his second paper on gas theory, submitted in December 1924, that Einstein highlighted this failure of statistical independence, even for an ideal gas of non-interacting particles; but still he did not make the connection with Gibbs, despite the fact that the generic phase also implies a failure of statistical independence, as needed to solve the Gibbs paradox.

We are back to the crux of the matter: how can this connection have been missed? It was not only Einstein; Schrödinger, who unlike Einstein continued to work on the new statistics throughout 1925, missed it too. But for either to make the connection something more was needed: the recognition that Einstein's great paper of 1905, 'On a heuristic point of view concerning the production and transformation of light', was in an important sense misleading. It purported to show that light in the Wien regime was made of statistically (or 'mutually')[10] independent particles, but the generic phase gives the same fluctuation formula, despite the failure of statistical independence. It was not a footnote they were missing, it was part of their footing.

If Einstein and Schrödinger did not remark on the connection, no wonder the young Dirac missed it too. Dirac went on to argue that states related by exchange of identical particles should be identified, just as had Gibbs, not only because no experiment could distinguish them, but because it seemed there was no element in the new formalism of quantum mechanics corresponding to such an interchange. The happy coincidence of that which was measurable with that which was expressible in matrix mechanics was for Dirac a great virtue of the new theory. This seemed to shut the door on any classical analogue of the quantum indistinguishability concept, for classically, there are always the trajectories. But here Dirac too was misled: quantum trajectories of non-interacting indistinguishable particles are quite easily defined.[11]

Since absent from all the historic papers on quantum statistics, it is hardly surprising that historians missed it subsequently. Gibbs, tragically, was unable to say more on the matter; he died unexpectedly in 1903, at the age of 64.

It seems there is no option, in substantiating these claims, but to revisit this history in detail. There are three important derivations of the Planck distribution at issue: by Planck in 1900, by Debye in 1910, and by Bose in 1924 (a fourth, the standard derivation after 1911, is also briefly considered). They carry the main line of the story. But we need to revisit Einstein's light quantum paper, and more, we need the 19th century background in classical statistical mechanics, above all concerning Gibbs' generic phase and its relation to Boltzmann's combinatoric definition of the entropy. We start with this.

---

[10] In Einstein (1905) it was 'mutually independent', in (1925) 'statistically independent', but Einstein made no comment on the shift in terminology and it is unlikely that it signified anything.
[11] See Saunders (2006, pp. 199–200), (2013 p. 358–359), Caulton (2011), Dieks and Lubberdink (2011 pp.1057-8).

## 2. The Classical Background

### 2.1 *Einstein's concept of 'mutual independence' and Gibbs' paradox*

We begin with Einstein's own stage-setting, in 1905, for his argument for light quanta. According to what he called 'the principle introduced into physics by Mr Boltzmann', the entropy of a system 'is a function of the probability of its state', denote $W$. If then two systems $A$ and $B$ do not interact with each other, the thermodynamic probability $W_{A+B}$ of the state of the composite should be the product of the thermodynamic probabilities of the states of each taken separately. Correspondingly, the total entropy should be the sum:

$$W_{A+B} = W_A . W_B \implies S_{A+B} = S_A + S_B.$$

This additivity of the entropy, for thermodynamic systems that are isolated from one another, is to this day a staple property of the entropy function however it is defined. From this Einstein deduced that for any entropy change:

$$S - S_0 = k \ln W^{rel}$$

where $W^{rel}$ is the 'relative probability' of the state with entropy S, with $S_0$ the initial entropy—the ratio of final and initial probability.[12]

Einstein's rational for additivity was that the instantaneous states of the two isolated systems $A$ and $B$ are, in his words, 'mutually independent events'. But he applied the concept more widely, not just to many-particle states of isolated thermodynamic systems, but to the one-particle states of particles that make up such systems. Thus if there are $Z$ one-particle states available to a gas of $N$ particles, degenerate with respect to the energy, and if the probability of finding a particle in any one of these states is $1/Z$, and the particles do not interact with each other, then the probability of finding all $N$ particles in a given one-particle state is $1/Z^N$ – so again, the probabilities factorise.

Einstein immediately put this factorizability condition to use:

> Let us consider a part of the volume $V_0$ of magnitude $V$ and let all $N$ moveable points be transferred into the volume $V$ without any other change in the system. It is obvious that this state has a different value of entropy $(S)$, and we now wish to determine the entropy difference with the aid of Boltzmann's principle.
> We ask: How great is the probability of the last-mentioned state relative to the original one? Or: How great is the probability that at a randomly chosen instant of time all $N$ independently movable points in a given volume $V_0$ will be contained (by chance) in volume $V$?
> Obviously, for this probability, which is a 'statistical probability',[13] one obtains the value

---

[12] Einstein used the same symbol $W$ for the probability of a state and for the relative probability connecting two states, but we shall try to keep them apart.

[13] Einstein had previously advocated an approach to the meaning of probability in physics based on sojourn times, and a weak notion of ergodicity (the fraction of time spent in a given state); this the 'statistical' definition.

$$W^{rel} = \left(\frac{V}{V_0}\right)^N ;$$

from this, by applying Boltzmann's principle, one obtains

$$S - S_0 = k \ln\left(\frac{V}{V_0}\right)^N. \qquad (p.96) \qquad (1)$$

The argument is simple and suggests that the relevant state space is the Cartesian product of one-particle phase spaces γ

$$\Gamma = \gamma \times \ldots \times \gamma \qquad (2)$$

with volume measure

$$W = \mathcal{V}^N \qquad (3)$$

where $\mathcal{V}$ is the volume (Lebesgue measure) of γ (or a region of γ). For a gas in equilibrium in spatial volume $V$, $\mathcal{V}$ as a dimensionless real number is of the form

$$\mathcal{V} = \frac{Vf(\sigma)}{\tau}.$$

where $f(\sigma)$ is a function of the intensive parameters and $\tau$ is a unit on γ with the dimensions of action. From Boltzmann's principle, the entropy is then:

$$S = k\ln(Vf(\sigma))^N - Nk\ln\tau. \qquad (4)$$

But just this leads to the Gibbs paradox.

The paradox is this. Let two samples of the same gas at the same temperature and pressure be confined to regions $A$ and $B$, separated by a partition, with available spatial volumes $V_A$ and $V_B$ and particle numbers $N_A$ and $N_B$. The initial entropy $S_{A+B}$ is the sum of the entropies $S_A$, $S_B$ of each taken separately; $S_{A\cup B}$ is the equilibrium entropy after the partition is removed. When the two samples mix, there should be no change in entropy (at least, no change in entropy that can be measured). Yet a simple calculation from (4) shows that the entropy increases:

$$S_{A\cup B} - S_{A+B} = kN_A \ln\frac{V_A + V_B}{V_A} + kN_B \ln\frac{V_A + V_B}{V_B}.$$

The associated thermodynamic probability $W_{A\cup B}$ of the two gases after mixing factorises

$$W_{A\cup B} = \mathcal{V}^N = \mathcal{V}^{N_A} \cdot \mathcal{V}^{N_B} = W_A \cdot W_B, \qquad (5)$$

as does that for the gases with the partition in place. The relative probabilities thus also factorise, as follows directly from (1), applied to gas $A$ and $B$ separately (each undergoing free expansion into volume $V$). The states of the two samples of gas are mutually independent, in Einstein's sense. But a non-zero entropy of mixing appears to be the inevitable consequence: in the case $V_A = V_B$, the increase on removing the partition is $kN \ln 2$, corresponding, for each particle, to a doubling in the number of accessible states in γ. It further follows that the entropy is not extensive: it does not scale with particle number and spatial volume, even for a homogeneous substance. So much is obvious from (3), but it follows from the non-zero entropy of mixing in the Gibbs set-up as well, for the gases in the two chambers at the moment the partition is removed just are parts of the gas as a whole. These are not the properties of the thermodynamic concept of entropy.

## 2.2 *The generic phase and Gibbs' paradox*

Gibbs's solution to these conundrums lay in the generic phase. He proposed that the correct phase space for *N* identical particles is not (2), but the space resulting from the identification of points in Γ representing the *N* particles that differ only in the permutation of particles – that differ only in which particle has which one-particle state in γ, as determined by sequence position in the Cartesian product. In more modern terms, he proposed that the correct phase-space structure for identical particles be the quotient space of Γ under the permutation group Π:

$$\tilde{\Gamma} = \gamma \times \ldots \times \gamma / \Pi.$$

To define this let $\langle q_k, p_k \rangle$ be coordinates (position and momentum) on the kth phase space γ in the Cartesian product (2), , $k = 1, \ldots, N$. Then a permutation $\pi$ of N particles has the action on Γ:

$$\pi: \langle \langle q_1, p_1 \rangle, \ldots \langle q_N, p_N \rangle \rangle \to \langle \langle q_{\pi(1)}, p_{\pi(1)} \rangle, \ldots \langle q_{\pi(N)}, p_{\pi(N)} \rangle \rangle.$$

Because they form a group, the relation on points of Γ related by some permutation $\pi \in \Pi$, is an equivalence relation. The equivalence classes under this relation are then the points of the quotient space $\tilde{\Gamma}$. In terms of coordinates on Γ, it is the same *N* pairs of positions and momenta, but unordered, rather than an *N*-tuple. If they are all distinct, a point in $\tilde{\Gamma}$ corresponds to *N*! distinct points in Γ, and the motion of a point in $\tilde{\Gamma}$ corresponds to *N*! distinct motions of points in Γ. They all describe the same *N* one-particle trajectories in γ; they differ only in which factor position each trajectory is assigned. There is no such redundancy for the generic phase.

As a transformation on Γ, the mapping $\pi$ seems no more than a renaming exercise, and hence an essentially trivial symmetry with only a passive interpretation, but there is a difference on passing to the quotient space. Not only is there a change in topology, but the beginning and end points of permutations *actively* interpreted must be identified. Smooth motions taking place over time, whereby two or more particles interchange their positions and momenta, can be described using either the specific phase or the generic phase: it is only that in the latter case the initial and final points of the generic phase space are one and the same. As the motion of a point in $\tilde{\Gamma}$, it is a closed loop, whereas in Γ (for each of the corresponding *N*! motions) it is an open curve. The *N*! open curves all describe the same *N* one-particle trajectories in γ, by which particles are actively interchanged. There is nothing *physically* puzzling about the generic phase, as describing those interweaving *N* trajectories without any redundancies, although it may introduce mathematical questions as to how that is to be achieved. There is every reason to think that this is exactly how Gibbs understood the matter.

To see that that the Gibbs paradox does not arise with the generic phase, suppose for simplicity that no two particles have exactly the same positions and momenta (such points anyway have Lebesgue measure zero). Then there are *N*! points in Γ for each point in $\tilde{\Gamma}$, so the volume measure on $\tilde{\Gamma}$ is:

$$\widetilde{W} = \frac{\mathcal{V}^N}{N!}. \tag{6}$$

The equilibrium entropy prior to mixing is by additivity (for the systems are initially isolated):

$$\tilde{S}_{A+B} = \tilde{S}_A + \tilde{S}_B = k \ln \frac{V_A^{N_A}}{N_A!} + k \ln \frac{V_B^{N_B}}{N_B!} + kN \ln f(\sigma) + N \ln \tau.$$

After the partition is removed and the particles are mixed it is:

$$\tilde{S}_{A \cup B} = k \ln \frac{(V_A + V_B)^{N_A+N_B}}{(N_A + N_B)!} + kN \ln f(\sigma) + N \ln \tau.$$

In the Sterling approximation

$$\ln x! \approx x \ln x - x$$

and given $N_A/V_A = N_B/V_B$ (as densities in the two chambers are the same) the difference in the two expressions vanishes. In the infinite limit (at constant density) the result is exact. There is no Gibbs paradox for the generic phase.

But (6) does not factorise. Correspondingly, after mixing, (5) is not satisfied:

$$\widetilde{W}_{A \cup B} = \frac{\mathcal{V}^{(N_A+N_B)}}{(N_A + N_B)!} \neq \frac{\mathcal{V}^{N_A}}{N_A!} \cdot \frac{\mathcal{V}^{N_B}}{N_B!} = \widetilde{W}_A \cdot \widetilde{W}_B \,. \tag{7}$$

The gases, when mixed, fail Einstein's criterion of mutual independence. Prior to mixing they are mutually independent, for they are separated by the partition, but not after. We come back to this in §4.2.

## 2.3 *Generic phase and Boltzmann's permutability*

We have considered a simple model of the Gibbs set-up, but Gibbs in the last chapter of his *Principles* treated the generic phase in general terms. He only discussed the Gibbs set-up at the very end, indeed, in the very last paragraph. He concentrated rather on showing how to define the generic phase in terms of the various ensembles (probability distributions) he had introduced in earlier chapters.

Commentators since have found this chapter cryptic,[14] the book as a whole overly mathematical. But to readers in the German-speaking world the cardinal difficulty was that it appeared to be wholly inconsistent with Boltzmann's combinatorial approach to the definition of thermodynamic probability, which clearly relied on the specific phase; a point on which Gibbs was completely silent.[15] Language as such was not the difficulty (the book appeared in German in 1904), but translation, still, was needed.

Boltzmann's concept of microstate ('complexion') was an assignment of particular particles to one-particle states. If the latter are enumerated $1, 2, \ldots, k, \ldots$, the coarser level of description or macrostate ('state distribution') is provided by the numbers $n_k$ ('occupation numbers') of particles in the kth state. Evidently there are

$$W = \frac{N!}{n_1! \, n_2! \ldots n_k! \ldots}$$

distinct complexions for each state distribution, corresponding to all possible exchanges of the $N$ particles between (but not within) the one-particle states (hence the need to divide by the $n_k!$'s). $W$ was called by Boltzmann the 'permutability' of the state distribution.

On normalising, Boltzmann assumed, the permutability is the probability; equivalently, he assumed complexions to be equiprobable. The equilibrium state distribution is then the most probable, consistent with fixed total particle number and total energy $E$. If $\varepsilon_k$ is the energy of the $k^{th}$ state, and if there are Z states in total, it is the extremal of the permutability consistent with the constraints:

$$\sum_{k=1}^{Z} n_k = N, \qquad \sum_{k=1}^{Z} \varepsilon_k n_k = E. \tag{8}$$

---

[14] Or worse: 'mystical' (Van Kampen 1984 p.309), 'the writings of an old man of rapidly failing health' (Jaynes 1992 p.2).
[15] In his *Principles* Gibbs referred to a number of Boltzmann's writings, but not to the 1877 memoire on the combinatorial approach to the entropy (although he surely knew it well).

Introducing undetermined Legendre multipliers for each of the constraint equations (the second to be identified as the temperature) Boltzmann had been able to derive, or rederive, some of his most important equations.

To all this, the difficulty posed by Gibbs' notion of generic phase is obvious. Complexions are specific phases; using generic phases instead, *the permutability is always unity*. A microstate (generic phase) only reports the number of particles in each one-particle state, not which particles are in which state. A state distribution just is a microstate. The idea makes nonsense of Boltzmann's combinatorial method – or so it seemed.

Still, we can say more about their connection. Consider first the relation of the permutability to the specific phase space measure (3). As before, let there be $Z$ coarse-grained one-particle states, but now degenerate with respect to the energy, each of them equiprobable. Then there are $Z^N$ available microstates, all of the same energy, all equiprobable. But equally, we can collect together those microstates with the same state distribution $\boldsymbol{n} = \{n_1, \ldots, n_Z\}$ ($W$ in all), and then sum the permutabilities. The result must be the same:

$$\sum_{\boldsymbol{n}; \sum_{k=1}^{Z} n_k = N} \frac{N!}{n_1! \, n_2! \ldots n_Z!} = Z^N. \tag{9}$$

Indeed, if the $n_k$'s are non-negative integers, (9) is an arithmetical identity.

There is an analogous combinatorial treatment for Gibbs' volume measure (6). Under the same coarse graining, and the same assumption of degeneracy of the energy, the total number of available microstates just is the total number of state distributions, but now without the weightings (the permutabilities) in (9). It is the number:

$$\sum_{\boldsymbol{n}; \sum_{k=1}^{Z} n_k = N} 1 = \frac{(N+Z-1)!}{(Z-1)! \, N!}. \tag{10}$$

For non-negative $n_k$'s, this too is an arithmetical identity. To obtain the connection with (6), observe that for $Z \gg N$:

$$\frac{(N+Z-1)!}{(Z-1)! \, N!} \approx \frac{Z^N}{N!} \tag{11}$$

in agreement with (6) for $\mathcal{V} \approx Z$.

The identity (10) is so important to our history that we do well to prove it explicitly. Observe that any ordered sequence of $Z - 1$ vertical strokes and $N$ crosses determines a unique microstate satisfying the constraints (8), a unique set of occupation numbers, and conversely.[16] There are $(N + Z - 1)!$ possible orderings of $N + Z - 1$ distinct symbols, but since only two are distinct we must divide by the number of permutations of the ×'s among themselves ($N!$ in all), and by the number of permutations of the |'s among themselves (($Z - 1$)! in all), yielding (10). Call this the 'symbol-sequence argument' for future reference.[17]

---

[16] Thus, to illustrate, for $Z = 7, N = 6$ the sequence ×× | × || ×× | × | determines the state distribution $n_1 = 2, n_2 = 1, n_3 = 0, n_4 = 2, n_5 = 1, n_6 = 0$, and conversely.

But the fact remains: every state distribution is equally probable. How, using Gibbs' generic phase, can Boltzmann's results be explained? Whether or not this simple difficulty was the reason Gibbs' concept of generic phase was found so baffling, there is no doubt that it *was* found baffling. Alas, Gibbs did not live to explain.

The key to reconciling the two is to draw a distinction between macrostate and microstate of a rather more robust kind – and one that is *also* to be found in Boltzmann's 1877 memoire. It figured in Gibbs' writings as well, and in the Ehrenfests' celebrated encyclopaedia article, 'The Conceptual Foundations of the Statistical Approach in Mechanics' of 1912, although in a singularly unhelpful form. The distinction is based on the idea not of a single coarse-graining of the one-particle phase space γ into regions degenerate with respect to the energy, but two: a further, finer-grained level of description of each region, in terms of *cells*. The former must be large enough to contain many particles, so that the Sterling approximation can be applied to the occupation numbers, but small enough so they all have approximately the same energy (this can always be arranged by increasing the spatial volume). They were called 'elementary regions' by Einstein in 1925, and we follow his usage (sometimes abbreviated to 'regions'). At the fine-grained level, however, Sterling's approximation is not needed, and the cells can be taken as small as desired ($\tau$ can be taken as small as desired – so that its size makes no difference to entropy differences).

Let the elementary regions be parameterised by $s = 1, 2, \ldots$, and let region $s$ contain $z_s$ fine-grained cells, each of equal phase space volume $\tau$ in γ. If $n_s$ particles are independently distributed over $z_s$ cells (specific phase), degenerate with respect to the energy, there is no further constraint on the microstate, so the number of accessible microstates is given by (9):

$$W_s = z_s^{n_s}. \tag{12}$$

Considering all regions s, and for a fixed partition of the total number $N = \sum_s n_s$ of particles into groups of $n_s$ particles, the total number of microstates is just the product of the $W_s$:

$$\prod_s z_s^{n_s}. \tag{13}$$

But there are many ways the $N$ particles may be partitioned into groups of $n_s$ particles – precisely Boltzmann's permutability, where now the elementary regions play the role previously played by cells. The total number of distinct microstates is therefore:

$$W = \frac{N!}{n_1! n_2! \ldots n_s! \ldots} \prod_s z_s^{n_s}. \tag{14}$$

If now we set up the same model using the generic phase, we replace (12) by (10) for each elementary region, and obtain instead of (13) the product over regions $s$

$$\widetilde{W} = \prod_s \frac{(z_s + n_s - 1)!}{(z_s - 1)! n_s!}. \tag{15}$$

---

[17] It first appeared in Ehrenfest and Kamerlingh-Onnes (1914). Ehrenfest wrote to Lorentz that he will laugh at the proof 'as at a joke', and that 'it was a lot of fun for us, and everyone who saw it' (quoted by Darrigol 1991 p.253); but he was unamused by those who took seriously the idea of the energy elements as things, as we shall see.

There is of course no further permutability factor, for the generic phase, as was introduced in (14), corresponding to permutations of particles between different regions, for particle permutations between different regions do not introduce any additional states, no more than they do within regions. If we now take the limit $z_s \gg n_s$, as classically we should, we obtain

$$\widetilde{W} \approx \prod_s \frac{z_s^{n_s}}{n_s!} . \qquad (16)$$

The result is identical, apart from the unwanted overall factor $N!$, with Boltzmann's permutability (14). The troublesome $N!$, spoiling extensivity, is removed. It had been frequently dropped by Boltzmann without any comment,[18] but here it has a principled rational. The standard procedure of maximising (14) to determine the equilibrium state applies to (16) just as well. Boltzmann's combinatorial method is entirely consistent with Gibbs' generic phase, differing only in the overall factor $N!$.

The apparent contradiction between the generic phase and Boltzmann's combinatorial method is thus fairly easily removed. But paradoxically, it became harder to make out after Planck's discovery of 1900, because his derivation of the radiation formula hinged on the use of Eq.(10) and on *not* taking the limit $z_s \gg n_s$. This combinatorics factor, whilst well-known to Boltzmann and to Gibbs, was following Planck's discovery imbued with a new and mysterious significance, in a context that seemed to have nothing to do with Gibbs paradox or the extensivity of the entropy. It became, along with Planck's constant, a signature of the quantum, alien to classical theory.

Whether for this reason, or by oversight, the correspondence just set out appears to have been missed by everyone working on quantum theory, meaning, by 1911, almost everyone. It was clearly missed by the Ehrenfests, for whilst they compared certain aspects of Gibbs' approach to Boltzmann's, they did not comment on the generic phase. They wrote down (14) but with the added (and unwarranted) assumption $z_s = \omega$ for all s, with the result that the correspondence between (14) and (16) was completely invisible.[19] Had they understood it they would not have thus concealed it. It was surely missed by Lorentz and the young Hugo Tetrode, in 1914. Tetrode had suggested that the division by $N!$ had been explained by Gibbs, a point that Lorentz found questionable. Certainly the needed correspondence Eq.(12)-(16) was not to be found in Gibbs' writings. Nor could the argument have been abbreviated; for example, it would have been self-defeating to divide the permutability Eq.(14) by $N!$ on the simple ground that the particles are exactly similar, so that permutations of complexions did not count, for that would remove the basis of introducing the permutabilities in the first place. Tetrode acquiesced in Lorentz's criticism[20], and even when the correspondence (12)-(15) was eventually brought out into the open, by Einstein in 1925, the crucial link, Eq.(16), was absent. The point still was not made.

There is another feature of this correspondence of great importance to our story. Eq.(15), (16) show two quite distinct senses in which the permutation of particles leaves the generic phase unchanged. One takes place *within* a given region, where the interchange of particles of approximately the same energies (momenta, location) among cells leave the microstate unchanged (hence the division by $n_s!$ for each s, here as in (6)). The other takes place *between* regions, where particles of very different energies (momenta, locations) are interchanged, that likewise do not produce a new microstate (hence the absence of the permutability). In the former case the particles involved have approximately the same properties; they are indistinguishable in the everyday, epistemic sense. But in the latter case the interchanged particles are from completely different regions of $\gamma$, and as such they have completely different positions and momenta. They are manifestly distinguishable, in the epistemic sense. That is: particles indistinguishable in Gibbs' sense, using the generic phase, may differ wildly in their state-*dependent* properties, as determined by different regions in $\gamma$; it is only their state-*independent* properties that must be the same.

---

[18] Ehrenfest and Trkal (1920) subsequently gave it a rational using Boltzmann's ideas ; but pace van Kampen (1984), whilst it offers a solution to the extensivity puzzle, it does not in itself solve the Gibbs paradox. Swendsen has recently worked it up in this way, but with a heavy dependence on probabilistic concepts (Swendsen 2018).

[19] As a result $\prod_s z_s^{n_s} = \omega^N$ in Eq.(14), a constant like $\tau$ that can be discarded (Ehrenfest and Ehrenfest, 1912 p.27).

[20] Declaring the reason for dividing the permutability by $N!$ to be 'a difficulty question'; see Darrigol (1991 p.276-7) for further discussion.

## 2.4 *Gibbs on indistinguishability*

Can we safely attribute this concept to Gibbs? He was not entirely explicit on the matter. The question of whether to use the specific or generic phase he thought a matter of 'practical convenience'. He settled on the generic phase, but went on to use his ensemble method to define probability distributions (and equilibrium distributions) using an ensemble of gasses each defined by a *specific* phase. The ensemble was an imaginary collective, the members mere 'creatures of the imagination', and as such

> The perfect similarity of several particles of a system will not in the least interfere with the identification of a particular particle in one case with a particular particle in another. (p.188).

This may have suggested to his readers that points of the *N*-particle generic phase space cannot even be *imagined* other than in terms of specific phases – that the generic phase of a gas can only mean a certain ensemble of gases, each with a definite specific phase (a probability distribution on specific phase space, as understood by Bach). But I suggest we should take Gibbs rather more at his word: it was a question of convenience.

We earlier gave a simple counting argument for the relation between the two volume measures (3) and (6) on $\Gamma$ and $\widetilde{\Gamma}$ respectively. For an analytical proof of the kind that Gibbs was familiar with, consider the simple case of a 1-dimensional interval of the reals $\lambda = [a, b]$ in place of $\gamma$. Let $\langle x_1, \ldots, x_N \rangle$ be Cartesian coordinates on $\mathbb{R}^N$, and hence on $\Lambda = \lambda \times \ldots \times \lambda$. The quotient space $\widetilde{\Lambda}$ under the permutation group for N particles $\Pi^N$ can then be defined as the set of points $\{\langle x_1, \ldots, x_N \rangle \; ; a \leq x_1 \leq \cdots \leq x_N \leq b\}$, so the ordering no longer specifies which particle has which coordinate.[21] There is only the trivial action of the permutation group on $\widetilde{\Lambda}$ (as the identity). The volume measure is:

$$\int_{\widetilde{\Lambda}} d\widetilde{\mu} \stackrel{\text{def}}{=} \int_a^b \int_a^{x_N} \ldots \int_a^{x_2} dx_1 \ldots dx_N = \frac{(b-a)^N}{N!}$$

(in agreement with Eq.(6)). Thus for $N = 2$, given two real numbers $x, x'$, $a \leq x \leq b$, $a \leq x' \leq b$, we write $x_1 = x, x_2 = x'$ if $x \leq x'$, and $x_1 = x', x_2 = x$ otherwise. No labelling of *particles* is involved (Figure 1).

---

[21] If coincidences are excluded, we use the simplex $a \leq x_1 < \cdots < x_N \leq b$ instead.

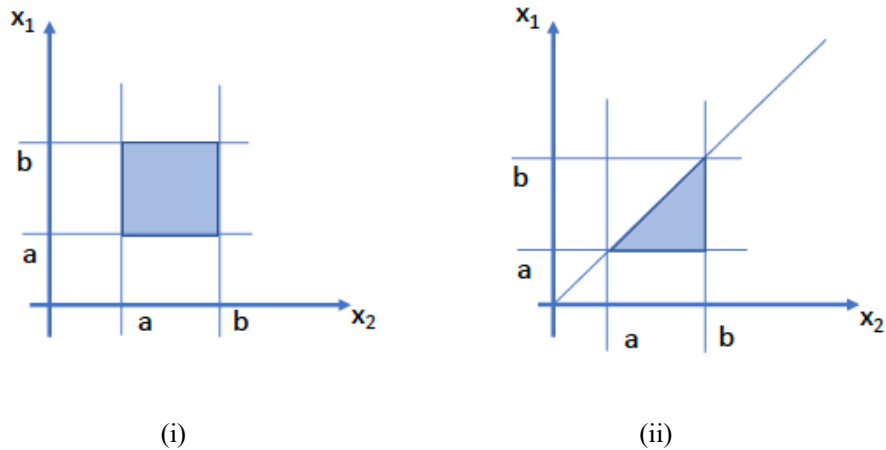

(i)             (ii)

Figure 1: 2-particle phase space. Coordinates for the shaded region in (i) consist of ordered pairs of real numbers (specific phase); for (ii), they are defined by unordered pairs of real numbers (generic phase).

The correspondence gives a simple ansatz for determining averages of functions on the quotient space, as for any completely symmetric function $f: \lambda^N \to \mathbb{R}$:

$$\int_{\tilde{\Lambda}} f d\tilde{\mu} = \frac{1}{N!} \int_a^b \int_a^b \ldots \int_a^b f(x_1, \ldots, x_N) dx_1 \ldots dx_N .$$

This is surely what Gibbs had in mind when he said, in favour of using the specific phase (and simply dividing by $N!$ at the end):

> For the analytical description of a specific phase is more simple than that of a generic phase. And it is a more simple matter to make a multiple integral extend over all possible specific phases than to make one extend without repetition over all possible generic phases. (p.188)

There can be no doubt that Gibbs advocated identifying phases that differed by permutations alone. When he did, finally, come to the mixing of like gases, he concluded:

> It is evident, therefore, that it is equilibrium with respect to generic phases, and not with respect to specific, with which we have to do in the evaluation of the entropy (p.207).

He used the term 'indistinguishable' only once, at the beginning of the chapter:[22]

> If two phases differ only in that certain entirely similar particles have changed places with one another, are they to be regarded as identical or different phases? If the particles are regarded as indistinguishable, it seems in accordance with the spirit of the statistical method to regard the phases as identical.

He almost certainly meant 'indistinguishable in their state-independent properties'. In his great memoire of 1875, 'On the Equilibrium of Heterogeneous Substances', Gibbs had spoken of microstates as indistinguishable ('undistinguishable') if they differ only by permutations. He then had the epistemic concept in mind: they did not differ in their 'sensible properties'. But in his final work it seems it was not the epistemic notion, for he toyed with the idea that generic phases might be defined so as to take no account of what he called 'equivalent positions', meaning states degenerate with respect to the energy, also sensibly the same. He decided against it: the *only* difference from specific phases lay in the absence of data as to which particle has which position and momentum.

---

[22] Gibbs (1902 p.187); this appears to be the first use of the term 'indistinguishable' in the context of permutation invariance. Quoting a passage from Planck (1901), Kuhn (1978 p.121, 286) translates 'indifferente' as 'indistinguishable', but in a different context, and 'nicht unterscheidbar' and 'nicht zu unterscheiden' are the standard German terms today. Natanson used 'unterschiedslos gleich' (indiscriminately alike) in his German publication, but 'undistinguishable' in the English version (1911), probably picked up from Gibbs' earlier writings.

To conclude, with respect to the question as posed, if we may not safely attribute our realist concept to Gibbs, it is certainly reasonable to read him in this way – and so might have his contemporaries. Alas, at least in the German-speaking world, Gibbs' concept of the generic phase was widely dismissed as conventionalist, or operationalist, or anti-realist, or idealist, or just frankly incoherent, to be contrasted with Boltzmann's realism.[23]

## 2.5 *Indistinguishability in Boltzmann's writings*

Ludwig Boltzmann, in his great study of 1877, has himself been attributed the concept of particle indistinguishability, for as a prelude to defining a coarse-graining (and indeed a fine-graining) of the one-particle state space $\gamma$ (into regions of position and momentum), he coarse-grained the total energy $E$, in an idealised model in which three-dimensional space was absent altogether. Given a unit $\varepsilon$ of energy, small in comparison to $E$, he defined what he called a 'geometric progression' $0, \varepsilon, 2\varepsilon, ..P\varepsilon$ with $P\varepsilon = E$; the members of this progression he called 'energy elements'. A microstate (complexion) is a specification of the energy element assigned each particle (allowing for repetitions), summing to $P\varepsilon$. A state distribution specifies the numbers of particles assigned each energy-element, and the equilibrium distribution is the one realised by the greatest number of microstates. Boltzmann imagined his complexions to be selected at random, using a ball and urn model of probability.[24]

In terms of our earlier remarks on Boltzmann's method, this model amounted to the choice of one-particle energies $\varepsilon_k = k\varepsilon$, with $k$ a non-negative integer, of total energy $E = P\varepsilon$. But as such it was easily translated into another model, in which there is a *multiplicity* of $P$ elements, each of identical magnitude $\varepsilon$ (and we shall use 'energy grade' henceforward for members of Boltzmann's progression $0, \varepsilon, 2\varepsilon, ..P\varepsilon$). A microstate can then be redescribed as the specification of the number of energy elements $\varepsilon$ assigned each particle *where there is no distinction as to which is assigned which particle*. It is only a redescription: they are still complexions, and the state distributions are the same, save that the enumeration begins with the zero (to indicate the state of zero energy) and terminates in $P$ (as the state of the greatest possible energy grade $P\varepsilon$). The permutability is:

$$\frac{N!}{n_0! n_2! \dots n_P!} \tag{17}$$

and the particle number and energy constraints are:

$$\sum_{k=0}^{P} n_k = N, \qquad \sum_{k=0}^{P} k n_k = P.$$

As before, microstates (complexions) must be equiprobable (or of equal phase-space volume) if relative numbers of microstates are to determine relative probabilities (or ratios of phase-space volumes). The total number of microstates then provides a normalisation constant. Boltzmann wrote down without comment the binomial coefficient:

$$\binom{N+P-1}{P} = \frac{(N+P-1)!}{(N-1)!\,P!}. \tag{18}$$

---

[23] See Darrigol (2018) for a systematic survey, to the same broad effect.
[24] As such the draws could not be independent, as Boltzmann noted, as the energy constraint has to be satisfied.

He almost certainly had worked out the third and most surprising of the arithmetical identities important to our story, the identity, for non-negative numbers $n_k$, of (18) with:

$$\sum_{\boldsymbol{n};\,\sum_{k=0}^{P} n_k=N,\ \sum_{k=1}^{P} k n_k=P} \frac{N!}{n_0!\,n_1!\ldots n_P!}. \tag{19}$$

It follows directly from Boltzmann's reasoning that (19) is the total number of microstates (complexions) of N (distinguishable) particles, with total energy $P\varepsilon$; and it *is* an arithmetical identity that this number is (18) (it was called 'the number of combinations with repetitions'[25]). But there is the simpler route: consider the microstate as a distribution of $P$ energy units $\varepsilon$ over the $N$ (distinguishable) particles, without regard to which energy unit is assigned which particle. The total number of microstates of this kind, energetically accessible, is just the number of microstates for $P$ (indistinguishable) units distributed over N (distinguishable) particles – or N (distinguishable) one-particle states – the only constraint being the total number $P$. The result is an instance of the identity (10), save for the interchange of the roles of particles and one-particle states:

$$\sum_{\boldsymbol{n};\,\sum_{k=1}^{N} n_k=P} 1 = \frac{(N+P-1)!}{(N-1)!\,P!}. \tag{20}$$

There is evidently a duality between the two descriptions: (distinguishable) particles assigned energy-grades, with microstates counted by (19), and (indistinguishable) energy elements assigned (distinguishable) particles or one-particle states, with microstates counted by (20). It was the shift made by Planck in 1900 when he wrote down the binomial (18), citing Boltzmann's 1877 paper.

In this way the indistinguishability concept has been traced to Boltzmann.[26] But understood in this way, it is the indistinguishability of the energy elements $\varepsilon$, entities that are exactly alike in *all* of their properties (basically, only energy). It is not the Gibbs concept, it is not even the epistemic concept; it might be called the *Planck* concept, as Planck was the first to talk in this way, albeit not for long. For Boltzmann, of course, the energy elements or grades were mere calculational devices, of no physical significance, for the unit $\varepsilon$, like $\tau$, was arbitrary.

How much of the foregoing was known to our protagonists, and when? Gibbs, probably, knew all of it. Boltzmann knew much, but he may have missed Gibbs' concept of generic phase, for his life too was cut short.[27] The two giants of statistical mechanics abruptly left the stage, just as the story of the quantum properly got going. Planck was well-versed in combinatorics formula, but his knowledge of Boltzmann's writings was, as Kuhn put it,

---

[25] For example, by Netto (1898 p.29), where it is listed as one of four 'standard formula' of the new field of combinatorics. See Kuhn (1978 pp.282-3) for further commentary.
[26] Bach (1990) defends this reading of Boltzmann; Monaldi (2009 p.385) cautions against it, as do Sharp and Matschinky (2015),
[27] Boltzmann was inactive in his later years, and took his own life in 1906.

'spotty'[28]; much more so of Gibbs'. Ehrenfest was critical of Gibbs' *Principles* and sought other solutions to the extensivity puzzle; he ignored the Gibbs paradox. So did Einstein, who having discovered a number of Gibbs' results independently, came late to the *Principles*. Peter Debye was an early champion of Gibbs' approach to statistical mechanics, and in 1910 he wrote down a symmetrised version of Eq.(15) for the total count of microstates of the radiation field; but he was unwilling to take energy quanta (as he called them) as things, and he missed the approximation (11), and with it the connection to generic phase. In other writings he cited Gibbs in justification for the division by $N!$, and on this point won a convert in Planck, but neither he nor Planck thought it related to the quantum or to black-body radiation.[29]

Gibbs left a precious legacy for the discoverers of the quantum, but it was not received. The history of discovery would have been quite different had it been.

### 3. Planck's Black-body Radiation Formula

*3.1: Planck's Discovery*

Planck's route to the black-body formula has been widely documented and we shall be brief. He modelled matter in thermal equilibrium with radiation at frequency $v$ in terms of a material harmonic oscillator, termed 'resonator', with natural frequency $v$. On the basis of an elaborate electrodynamical argument Planck concluded that the mean oscillator energy $U$ must be related to the equilibrium radiation energy density $\rho$ as:

$$\rho = \frac{8\pi v^2}{c^3} U. \qquad (21)$$

Wilhelm Wien had some years earlier deduced certain constraints on $\rho$ on general thermodynamical grounds, and had proposed the simplest candidate with bounded energy, *the Wien black-body distribution*:

$$\rho = \alpha v^3 e^{-\beta v/T}. \qquad (22)$$

This introduced two new dimensional constants $\alpha$ and $\beta$, independent of frequency; it was this that excited Planck's interest. But although initially successful, it was clear by the summer of 1900 that Wien's law failed at high temperatures and low frequencies (the Rayleigh-Jeans regime), where $\rho$ appeared to depend linearly on the temperature. Using the same combination of simplicity and guesswork that had guided Wien, Planck proposed the new formula, *the Planck black-body distribution*:

$$\rho = \frac{8\pi v^2}{c^3} \frac{hv}{e^{hv/kT} - 1}.$$

For $hv \gg kT$, the Planck distribution goes over to the Wien distribution. Planck had earlier tried to derive the Wien law from first principles (and, embarrassingly, had claimed to succeed), and now set out to derive his new distribution law. From the first law of thermodynamics

$$dU = TdS + pdV$$

---

[28] Kuhn (1978 p.100) observes that Planck had prepared the thirteenth lecture of Kirchhoff's *Theory of Heat* for posthumous publication in 1894, which, unusually for the time, contained a detailed combinatorial derivation of Maxwell's velocity distribution.
[29] Debye (1910a), Planck (1915, 1921).

the entropy S of the resonators must satisfy the fundamental relation:

$$\left.\frac{dS}{dU}\right|_{V=const} = \frac{1}{T}. \tag{23}$$

Using Planck's distribution law to express 1/T first as a function of $\rho$, and then from (21) as a function of $U$, Eq.(23) is readily integrated to give the entropy of the resonator:[30]

$$S = k\left(\frac{U}{h\nu} + 1\right)\ln\left(\frac{U}{h\nu} + 1\right) - k\frac{U}{h\nu}\ln\frac{U}{h\nu}. \tag{24}$$

Conversely, from (21), (23), and (24), the Planck distribution followed. The latter was holding up to experiment beautifully, the other equations followed from well-known physical principles; Planck had only to justify Eq.(24).

There followed, said Planck in his Nobel prize lecture in 1918, 'the most strenuous work of my life'. We know he was eventually led to Boltzmann's 1877 memoire, for he cited it. Whether he there first found Boltzmann's binomial coefficient for the total number of microstates, or elsewhere – or, whether, indeed, as Kuhn has suggested, he worked directly from (24) – is disputed.[31] On writing down the expression (18), he noted that in the Sterling approximation its logarithm multiplied by Boltzmann's constant is:

$$k\ln\frac{(N+P-1)!}{(N-1)!\,P!} \approx Nk\left(\left(\frac{P}{N} + 1\right)\ln\left(\frac{P}{N} + 1\right) - \frac{P}{N}\ln\frac{P}{N}\right). \tag{25}$$

The similarity with (24) leaps to the eye. Suppose, indeed, there are N resonators, all with the same entropy (24), and suppose that the total entropy is the sum of the individual entropies. Then the entropy of the resonators is:

$$Nk\left(\left(\frac{U}{h\nu} + 1\right)\ln\left(\frac{U}{h\nu} + 1\right) - \frac{U}{h\nu}\ln\frac{U}{h\nu}\right)$$

an expression identical to (25) if:

$$\frac{P}{N} = \frac{U}{h\nu}$$

that is, if $P = UN/h\nu = E/h\nu$, where $E$ is the energy of the $N$ resonators. Planck had only to choose in place of Boltzmann's (arbitrary) unit $\varepsilon$ the quantum of energy $h\nu$.

Planck had the answer to a combinatorial problem, namely the Boltzmann binomial; he had only to find the problem. Here it is, in Planck's words:

> If $E$ is considered as an infinitely divisible quantity, the distribution can be made in an infinite number of ways. However, we consider– and this is the most important point of the whole calculation – $E$ as being composed of a completely definite number of finite equal parts, and make use for that purpose the natural constant $h = 6.55 \times 10^{-27}$ erg sec. This constant, when multiplied by the common frequency of the resonators, yields the energy element $h\nu$ in ergs; and by dividing $E$ by $h\nu$, we obtain $P$, the number of energy elements which have to be distributed among the $N$ resonators. (Planck 1900 239-40).

---

[30] This follows Kuhn's now widely-accepted reconstruction of Planck's route to his discovery.
[31] Mehra and Rechenberg (1982 p. 50) trace it to Boltzmann, and give arguments against Kuhn's suggestion.

The number of ways in which this can be done is given by Boltzmann's binomial (18) (although Planck gave no proof). From this and Boltzmann's principle, Eq.(25) follows, and with it the Planck distribution.

As we saw, Boltzmann in 1877 had spoken of an 'energy element' (energy-grade) as an integral multiple of the unit, the energy assigned a particle, a member of the progression $0, \varepsilon, 2\varepsilon, \ldots, P\varepsilon$, with $P = E/\varepsilon$. Planck used ' energy element' to mean the unit itself, and the plural to mean a multitude of units -- he spoke of the energy 'as composed of a definite number of finite equal parts'. He explains the expression (18) in terms of the number of distinct 'complexions' (microstates), borrowing Boltzmann's term for what he is careful to call a 'similar concept'. We have previewed these shifts already. They were slight and easily made, but they carried the suggestion that the energy elements $\varepsilon$ might be thought of as objects in their own right, and indeed as exactly indistinguishable from each other, not just indistinguishable in the epistemic sense.

This was not a picture that Planck especially liked or even intended,[32] and in his *Gas Lectures* of 1906 he offered another (that we need not consider here). There was much else that was puzzling. He had hitherto been content with a single resonator in each frequency range, but now invoked large numbers (needed to ensure the Sterling approximation was valid), without explanation. More confusingly, Planck used for the count of microstates what was for Boltzmann an overall normalisation factor for converting permutabilities $W$ to genuine probabilities. Correlatively, Planck did not seem to be extremalising anything: in what way was he deriving the *equilibrium* distribution for his resonators? He did allude to the more coarse-grained macrostate, whereby different energies were to be assigned to each 'group' of resonators (for each frequency range), with the total number of microstates given by the product of the W's (for each group); in this way he hinted at an expression of the form (15) (but did not write it down). His *Annalen* derivation submitted two months later differed from this sketch in significant respects.[33] Planck had made the discovery of the century, but he had no idea what it meant.

## 3.2 The Debye Model

The following decade saw repeated attempts to place Planck's black-body law on a sounder footing, including by Planck. The quantisation of energy, in units $h\nu$, featured in almost all of them.

By the time of the Solvay Conference of 1911, the first international conference on the new quantum theory, it seemed that Planck's law could be placed on a satisfactory footing in two quite distinct ways, both of which traced the quantum discontinuity to the unknown laws of atoms and molecules, impacting, through energy transfers with matter, only derivatively on radiation. The first was due to Debye, dispensing with Planck's resonator equation, modelling the radiation field directly. Normal modes of the electrodynamic field in a given spatial volume replaced Planck's material resonators (and were just as distinguishable). The second was due to several authors but was first sketched by Planck[34] and Einstein in 1905 and 1906, using Planck's resonator equation, and in more detail by Lorentz in 1909. By 1911 this had become something of an orthodoxy.[35] In this approach only *material* harmonic oscillators figured (and again were perfectly distinguishable). In neither approach were the energy quanta regarded as objects in their own right.

---

[32] As emphasised by Kuhn (1978), who makes a persuasive case that despite Planck's language, he did not even think the energies of individual resonators were quantised until about 1909. Planck's phrasing at this time may only have reflected the easier combinatorial route to the Boltzmann expression (with some version of the symbol-sequence derivation), rather than the cumbersome 'combination with repetitions' route, also in keeping with Kuhn (1978 p.127-30).

[33] There Planck considered only resonators in a single frequency range, and assumed equilibrium was already established (with the entropy given, as before, by the logarithm of the Boltzmann expression). From the Wien displacement law he then deduced $P = E/h\nu$.

Debye formulated his guiding idea in terms of what he called the 'elementary quantum hypothesis':

> The …major difference [with conventional gas theory] is the application of the elementary quantum hypothesis, which is analogous to that of Liouville's theorem. Just as the latter determines the transfer of energy from one degree of freedom to another in collisions, the former hypothesis in radiation theory makes it possible to ascertain the corresponding amount of energy transferred from one wavelength to another, provided it is caused by a material body.[36]

He proposed to make use of nothing else – and in particular, to make no use of Planck's resonator equation (21), based as it was on Lorentz's electrodynamics.

As was then well-known, if the electromagnetic field is treated as a collection of monochromatic waves or modes, subject to simple boundary conditions for volume $V$, the number of modes in the frequency range $[v, v + \Delta v]$ is:

$$n\Delta v = \frac{8\pi v^2}{c^3} V \Delta v \,. \tag{26}$$

In 1902 James Jeans had shown that this number is independent of the shape of the cavity. Debye called the collection of normal modes for volume $V$ and frequencies in the range $[v, v + \Delta v]$ the 'Jeans cube'. On the basis of his quantum hypothesis, he supposed the energies of the modes in the cube to be integral multiples of $hv$, what he called 'quanta'. He defined a macrostate as a function $f(v)$, specifying the number of quanta for each mode of frequency $v$; the combinatorics problem was to calculate the number of ways of distributing $f(v)n(v)\Delta v \,(= P)$ quanta over $n(v)\Delta v \,(= N)$ normal modes of the Jeans cube. Referring to Planck (1900), but with no other comment, he wrote down the answer as:

$$W_v = \frac{(N+P)!}{N!\, P!} = \frac{(n\Delta v + fn\Delta v)!}{(n\Delta v)!\,(fn\Delta v)!} \tag{27}$$

and for the total number of microstates (compare Eq.15):

$$W = \prod_v \frac{(n\Delta v + fn\Delta v)!}{(n\Delta v)!\,(fn\Delta v)!} \,. \tag{28}$$

---

[34] Planck went on to modify the theory to allow continuous energy absorption, but discontinuous emission (his so-called 'second theory'), an approach that eventually led to half-integral values of the energy (and a zero-point energy). It was this that appeared in the second edition of his *Lectures*.

[35] Although based on Planck's resonator equation, and hence on Lorentz's electrodynamics, it was thought it may still have a statistical validity. Not by Einstein, who pointed out that it required a mean resonator energy (of order $kT$) much greater than the energy unit, whereas in the Wien limit it was much smaller (1906 p.198).

[36] Debye (1910b p.1431); Einstein might have stated the quantum hypothesis similarly, but without the final proviso.

This was to be maximised to determine the equilibrium macrostate $\bar{f}(v)$, subject to the energy constraint:

$$E = \sum_v hv f n \Delta v.$$

Replacing the sum by the integral and substituting for $n\Delta v$ from (26) it follows:

$$E = \frac{8\pi h}{c^3} \int_0^\infty v^3 f dv.$$

If the logarithm of W is a maximum under variation of $f$, it must satisfy the equation:

$$\ln(1+\bar{f}) - \ln\bar{f} = \beta hv$$

where $\beta$ is the undetermined Lagrange multiplier for the energy constraint. The entropy for the equilibrium macrostate must satisfy the analogue of Eq.(8), fixing $\beta$ as $1/kT$. Therefore:

$$\bar{f}(v) = \frac{1}{e^{hv/kT}-1}$$

and the Planck distribution follows immediately from Eq.(26).

The crucial expression is Eq.(28), a product of symmetrised Bolzmann binomials (18) (replacing $N$-1 by $N$), one for each frequency range. Writing $n_v = f(v)z(v)\Delta v$ as the number of quanta in the frequency range $[v, v+\Delta v]$, and $z_v = n(v)\Delta v$ as the number of modes, or 'receptacles', we can rewrite (28) as:

$$W = \prod_v \frac{(n_v + z_v)!}{n_v! z_v!}$$

whereupon the comparison with Eq.(15) is transparent. There are no additional microstates corresponding to the interchange of quanta *between* different frequency ranges, but then it occurred to no-one, at this time, that quanta were things that could change continuously in time, so as to interchange their properties. The modes of the field were the physical entities for Debye. Bose was to invert this, and take the quanta as the physical things, and the modes of the field as one-particle states. For the Jeans cube, these states are approximately degenerate with respect to the energy, and the derivation of (15) applies immediately.

### 3.3 The Solvay Model

The third derivation of the black-body formula that we shall consider prior to Bose's used a different strategy to achieve the same end: to confine the quantum discontinuity to the unknown dynamics of molecules, preserving Maxwell's equations unchanged. The beauty of this approach was that save for the quantisation of the energy of the molecules (Planck's resonators) its credentials were impeccable; it was clearly based on Boltzmann's principles, whilst avoiding the troublesome Boltzmann binomial, Eq.(18). It remains standard pedagogy to this day. For brevity we shall call it 'the Solvay model'.

Let there be $N$ (distinguishable) objects, and let a state distribution as before be defined by the number $n_k$ of objects with energy $k\varepsilon$, with total energy $E = P\varepsilon$, subject to the constraints (8) with $\varepsilon_k = k\varepsilon$. A microstate is the specification of the energy of each object (one of the elements $\{0, \varepsilon, 2\varepsilon, \ldots, P\varepsilon\}$). The permutability is as before, Eq.(17). Assuming, as usual, the validity of the Sterling approximation, the entropy is:

$$k \ln W = Nk \ln N - k \sum_k n_k \ln n_k.$$

It is a maximum if it vanishes under variation in the $n_k$'s as constrained by (8). Introducing the usual Lagrange undetermined multipliers $\alpha, \beta$ (one for each constraint):

$$\delta \ln \overline{W} = -\sum_k (\ln \bar{n}_k + 1 + \alpha + \beta k \varepsilon) \, \delta n_k = 0$$

where now the variations $\delta n_k$ can be treated as independent, it follows:

$$\ln \bar{n}_k + 1 + \alpha + \beta k \varepsilon = 0.$$

On rearranging we obtain (absorbing $\alpha$ into a normalisation constant $A$):

$$\bar{n}_k = A e^{-\beta k \varepsilon} \tag{29}$$

where $A$ and $\beta$ are to be determined from the constraints (8) and the fundamental relation (23) (replacing $U$ by $E$). *Every single step* in this derivation had been given by Boltzmann, writing down formulae identical to the above in his memoire, more than three decades before, up to notational variations. Eq.(29) is known as the 'Boltzmann probability distribution'.

Supposing now, as Boltzmann did not, that the size of the unit $\varepsilon$ has a physical meaning and that the continuum limit should not be taken. Then the average energy of the objects is

$$U = \frac{\sum_k k \varepsilon \bar{n}_k}{\sum_k \bar{n}_k} = \frac{\varepsilon}{e^{\beta \varepsilon} - 1} . \tag{30}$$

If instead the continuum limit is taken, we obtain:

$$U = \frac{\int \epsilon e^{-\beta \epsilon} d\epsilon}{\int e^{-\beta \epsilon} d\epsilon} = \frac{1}{\beta}. \tag{31}$$

As usual, in both cases, $\beta$ is identified with $1/kT$ from the fundamental relation Eq.(23).

Suppose now that the objects are (distinguishable) Planckian resonators with natural frequency $v$, and that the energy unit is $\varepsilon = hv$. With these substitutions in (30), and from the resonator equation (21), the Planck distribution follows immediately. Working from the continuum limit (31) instead we obtain the Rayleigh-Jeans distribution:

$$\rho(v, T) = \frac{8\pi v^2}{c^3} kT .$$

The latter follows even more simply from the equipartition theorem applied to the normal modes of the Jeans cube, and from the Jeans number Eq.(26). As a *general* result, of course, it would be disastrous: the energy of radiation in any finite volume, at non-zero temperature, would be infinite, for the distribution diverges with frequency (the 'ultra-violet catastrophe' as Ehrenfest memorably called it). It is the discretisation of the energy of the resonators that makes the difference: at high frequencies, for $hv \gg kT$, resonators will rarely or never be excited by random motions at thermal energies of order $kT$, for the energy transfers needed will rarely or never be so large.

The result was rigorous (with the imprimatur of both Lorentz and Poincaré), and proved that discretisation of the energy was essential to the quantum theory of material oscillators. The implications transcended the black body radiation formula, and following Bohr's model of the atom, led directly to the Bohr-Sommerfeld quantisation rules. It was the locomotive driving to Heisenberg's matrix mechanics, but on the indistinguishability concept it had nothing to say. And there for more than a decade the matter lay.

## 4. Light Quanta

### *4.1 Natanson*

Ladislas Natanson, Professor of theoretical physics at the University of Cracow, had a long and distinguished career, but it is for his essay 'On the statistical theory of radiation' published in 1911 that he has been thought important by historians. It is widely accredited to be the first significant writing on the concept of quantum indistinguishability,[37] so it deserves special consideration. Although it seems it went unremarked, it may have been read. It was certainly accessible, as published in English (in two journals) and in German.

It began with the words:

> It is to Planck that we owe the fundamental conception on which the theory in its present state essentially depends. As is well known, Planck supposes that energy is not divisible without limit, but is constituted of an aggregate of discrete elements or units; and that the ultimate particles of which matter consists are capable only of absorbing, containing and emitting amounts of energy which are multiples of these finite and determinate units.

Natanson took Planck's early talk of energy elements literally, as a plurality of units $\varepsilon$ (Natanson used the symbol $e$). They were to be distributed over these 'ultimate particles of which matter consists' – what he went he went on to call 'independent material entities' (and then 'elementary receptacles'; 'independent' was not used again). A state distribution is now an assignment of energy elements to receptacles. These he initially proposed to treat in an even-handed way:

> When we speak of state-distributions we do not in any sense imply we can identify either receptacles or energy-units. To specify a definite state-distribution, we require to know the number of elementary receptacles to which any given number of units of energy has been allotted; but we are not bound to have any control over the receptacles taken individually, nor to be able to detect each energy-unit and to trace it through its course. The circumstances of the problem are quite different if we suppose that the elementary receptacles of which our problem consists *are* distinguishable; that we can lay hold of each receptacle and ascertain its identity.' (p.135)

In that case, we should use Boltzmann's term 'complexions', and indeed count them as equiprobable – but only so long as the energy-units are not distinguishable:

> In speaking as above of complexions, it is of course implied that we are capable of dealing with the individual receptacles; but the elements or units of energy are all regarded as being undistinguishably alike.' (p.136).

---

[37] Hund suggests Natanson's method 'is precisely that later used by Bose' (Hund 1974 p.51). Kastler (1983) and Bergia (1987) concur with Hund and are puzzled his paper received so little notice. According to Darrigol (1991 p.239), Natanson was the first to introduce the concept of indistinguishable particles in what he calls the 'molecular' approach to probability (in contrast to Gibbs' 'holistic' approach).

As Natanson goes on to argue, if the energy-units were distinguishable, the relevant microstates would be quite different; and if the latter were equiprobable, the former could not be: nature must decide. And indeed, nature has already chosen: 'and it may be recalled here that in every known demonstration of Planck's formula the same [indistinguishable] principle is employed'.[38]

Was Natanson right to say this? It is true that all the then-known derivations *could be* interpreted in terms of indistinguishable energy-units, or very like, and as we have seen, and as Natanson rightly insists, invariance of microstates under permutations of energy-units among receptacles (Planckian resonators, or modes of the Jeans cube) *is* then required. But by 1911 neither Planck, nor, at that time, virtually anyone (including Einstein) wished to interpret the energy distribution over resonators in terms of an assignment of energy-units over receptacles, be they resonators or modes of the electromagnetic field. Debye chose not to; Planck preferred, where possible, not to talk about radiation at all. On the Solvay model, resonators were assigned energy-grades, not energy-units. And those who took light quanta seriously, most notably Einstein and Ehrenfest, thought stronger arguments were needed to infer a particle structure to radiation, arguments that appeared to be restricted to the Wien regime. On all these points Natanson was silent.

What, at bottom, was Natanson's conception of indistinguishables? He nowhere explicitly identified his energy-units as light quanta. They were abstract from the beginning (he did not so much as mention the concept of frequency). That would seem to suggest they must have exactly the same energy (and since they have no other attributes, they must be exactly similar) – the Planck notion. His talk of 'laying hold of' and 'detection', on the other hand, suggests the epistemic notion. At any rate, it explicitly rules out entities that can be 'traced through their course', so they cannot possibly have trajectories. It is not the Gibbs concept.

Natanson introduced needed clarity. He set out some of the combinatorial expressions of §2.5, including the identity of (18) and (19) (but so had Boltzmann). He gave a derivation, not of the Planck distribution, but of the Boltzmann probability distribution, Eq.(29) – precisely following Boltzmann! His subsequent discussion of the difference between systems 'abundantly bestowed' with energy-units, and those 'poorly endowed', missing as he did the coarse-graining into frequency intervals, was flawed. He made no comment on statistical independence, or the lack of it. There may be no great puzzle here as to why his essay received so little commentary.[39]

Yet it may well have prompted the complaint against those who took 'energy elements' seriously, made by Ehrenfest and Kamerlingh-Onnes the following year, as appended to a note on the derivation of the Boltzmann binomial Eq.(18) (the symbol-sequence argument). They did not mention Natanson by name, but they surely had him in mind when they said:

> Planck does not deal with really mutually free quanta $\varepsilon$, the resolution of the multiples of $\varepsilon$ into separate elements $\varepsilon$, which is essential in his [original] method, and the introduction of these separate elements have to be taken 'cum grano salis'; it is simply a formal device entirely analogous to our permutations of the elements [symbols] $\varepsilon$ or 0 [in the symbol-sequence argument]. The real *object which is counted* remains the number of all the different distributions of N resonators over the energy-grades 0, $\varepsilon$, 2$\varepsilon$, with a given total energy $P\varepsilon$. ..... Planck's *formal device* (distribution of P energy elements ε over N resonators) *cannot be interpreted in the sense of Einstein's light-quanta* (1912 p.357, emphasis original).

---

[38] (Ibid p.139). Natanson continued 'The moment we suppose that individual energy-units are not altogether beyond our power of discerning them, the rule laid down by Planck, and adopted by nearly all writers who have treated of the subject, ceases to be applicable.' (ibid p.139).

[39] See Bergia (1987 p.234-5) for further discussion.

The key phrase is 'really mutually free quanta'. Planck's energy elements, whatever they were, could not be mutually free, except in a special regime. It is time to see why.

## *4.2 Einstein's Light Quantum Argument*

Einstein's 1905 argument for a particulate structure to radiation is a thing of beauty, but his wording in stating his conclusion was unfortunate:

> Monochromatic radiation of low density (within the range of validity of Wien's radiation formula) behaves thermodynamically as if it consisted of mutually independent energy quanta of magnitude $h\nu$.

There is no mention of locality, although this too was used in deriving his fluctuation formula – and was in fact the more essential assumption. As we shall see, despite the particular kind of mutual dependencies involved using the generic phase, the same fluctuation formula follows. Einstein's argument does not discriminate between the generic and the specific phase.

There are a number of similarities to Planck's derivation of his black-body formula, as outlined in §3.1. Like Planck, Einstein inferred the equilibrium entropy of radiation from the spectral density formula, and like Planck, he interpreted this entropy in terms of probabilities. The difference was that Einstein worked from the *Wien* distribution, Eq.(22), restricting attention to the high-frequency, low-temperature regime. Another difference is that he did not seek to *justify* the spectral law (he already knew the Wien distribution held only in this regime); he sought rather to learn from it. He began his paper with the observation that the equipartition theorem, applied to the classical electromagnetic field, gave the right behaviour at long wavelengths and high temperatures (what he called 'high energy densities', in view of Stefan's law), but not at high frequencies and low temperatures; he wanted to know why radiation in the Wien regime behaved so differently.

Einstein's starting point was the fundamental relation (23), but written in terms of densities, what he called 'the law of black bodies':

$$\left.\frac{d\varphi}{d\rho}\right|_{V=const} = \frac{1}{T}.$$

From the Wien distribution law (22), the right-hand side can be written as a function of $\rho$; on integrating:

$$\varphi = -\frac{\rho}{\beta\nu}\left(\ln\frac{\rho}{\alpha\nu^3} - 1\right). \tag{32}$$

Planck had found this equation in 1898 in exactly the same way; he used the same method to derive Eq.(24) from the Planck distribution two years later. Einstein's originality lay in what came next.

Einstein supposed that the radiation field, like a gas, may be subject to *fluctuations*. Specifically, he considered a fluctuation whereby the energy $E$ of radiation initially in spatial volume $V_0$ is subsequently located in a sub-volume $V$. The energy density thus changes from $\rho = E/V_0$ to $\rho' = E/V$ and with it the entropy density $\varphi$. From Eq.(32) the initial entropy is:[40]

$$S_0 = \varphi V_0 = -\frac{E}{\beta v}\left(\ln\frac{E}{\alpha v^3 V_0} - 1\right)$$

and after the fluctuation to volume V it is:

$$S = \varphi' V = -\frac{E}{\beta v}\left(\ln\frac{E}{\alpha v^3 V} - 1\right).$$

The difference is:

$$S - S_0 = \frac{E}{\beta v}\ln\frac{V}{V_0} = k\ln\left(\frac{V}{V_0}\right)^{E/k\beta v}.$$

But this has the same dependence on the two volumes as obtained for an ideal gas of N particles, Eq.(1), if only the exponent of the ratio of volumes can be interpreted as the number of particles – if only radiation consisted of 'independent, movable points'. In this way he deduced:

$$N = \frac{E}{k\beta v}$$

or, in Planck's notation, if $N = E/hv$. That is, if light were made up of mutually independent, moveable points, of number $E/hv$, then the fluctuation formula would be in accordance with that following from the Wien distribution. Locality is essential to the argument.[41]

But is mutual independence? If the volume measure on phase space is of the form $W \propto V^N$, the relative probability of the fluctuation is, according to Einstein, the ratio

$$\frac{W}{W_0} = \left(\frac{V}{V_0}\right)^N. \qquad (33)$$

This assumes the specific phase. But the ratio using Gibbs' generic phase is given by the same expression:

$$\frac{\widetilde{W}}{\widetilde{W}_0} = \left(\frac{V}{V_0}\right)^N$$

despite the fact that $\widetilde{W} \propto V^N/N!$ cannot be factorised.

Evidently we should read Einstein as requiring mutual independence – the factorisation condition – as holding only of the relative probabilities, so that only ratios in the probabilities need factorise. And so they do: each particle, using the Gibbs phase, has a relative probability $V/V_0$ of being found in $V$ if initially in $V_0$, and the total relative probability of all N particles initially in V being subsequently found in $V_0$, is the product of the relative probabilities for each of them, the same for the generic as for the specific phase. However, there are other fluctuations possible which do not factorise in this way – most notably, as in the Gibbs

---

[40] This step implicitly assumes extensivity of the entropy, as Einstein must have realised.
[41] Whether the 'movable points' must really be *points* will depend, presumably, on the precision with which the probability of the fluctuation is defined (perhaps only in the limit $v \to \infty$). Dorling (1971) argues the particles must be point-like (recently endorsed by Norton (2006)).

paradox.

Thus, consider the gas in volume $V_0$ as consisting of two gases, of particle number $N_A$ and $N_B$, each initially in volume $V_0$; each of which undergoes a fluctuation, the one resulting in the $N_A$ particles being localised in A, a sub-volume of $V_0$ with volume $V_A$, and the other resulting in the localisation of the $N_B$ particles in the disjoint region B, of volume $V_B$. If the two fluctuations are mutually independent, the relative probability of the two taken jointly is the product of the relative probabilities of each fluctuation taken independently, i.e. it is the product:

$$\frac{V_A{}^{N_A}/N_A!}{V_0{}^{N_A}/N_A!} \cdot \frac{V_B{}^{N_B}/N_B!}{V_0{}^{N_B}/N_B!} = \left(\frac{V_A}{V_0}\right)^{N_A} \cdot \left(\frac{V_B}{V_0}\right)^{N_B}. \tag{34}$$

But taken jointly, the relative joint probability is the ratio of the joint final probability to the joint initial probability:

$$\frac{V_A{}^{N_A}/N_A! \cdot V_B{}^{N_B}/N_B!}{V_0^N/N!}. \tag{35}$$

The two are not the same, and (35) does not factorise, no more than does (7). Eq.(34) is the same as using the specific phase, and leads to the Gibbs paradox, as we have seen. When the densities of particles are the same, $N_A/V_A = N_B/V_B = N/V$, the joint fluctuation is just the time inverse of the mixing process in the Gibbs set-up; on the generic phase, it yields no change in entropy. No more does the entropy density derived from the Wien distribution by (32) change for the joint fluctuation, for the joint energy and number density before and after are the same.

Whether probabilities or relative probabilities, if calculated by the generic phase, they do not factorise. Collections of indistinguishable particles in Gibbs' sense are not mutually independent of each other, in Einstein's sense. Of course that means that collections of *photons* are not mutually independent of each other, either, in Einstein's sense, not even in the Wien regime; for the Wien regime just is the classical limit of a gas of photons, which just is Gibbs generic phase, by Eq.(11).

What then is showing up away from the Wien regime, for indistinguishable particles, if not the failure of statistical independence? But failure come in degrees: away from the Wien regime, Eq.(11) fails as well as factorizability. Statistical dependencies, which hitherto were showing up only for collections of particles, now show up at the level of individual particles. The probability of an already occupied state is greater than that calculated using the specific phase, if every state distribution is equiprobable, on the generic phase.

Einstein had a more intuitive way of putting it. Away from the Wien regime, *wavelike* behaviour begins to show. After all, it was in the long wavelength, high temperature regime that classical electromagnetic waves behaved thermally as they should, obeying the equipartition theorem. He was further buttressed in this view by a second fluctuation formula for radiation, that he obtained in 1909, this time concerning the mean square deviation in energy in a given volume. It consisted in the sum of two terms, one characteristic of a gas of particles, the other of classical radiation. It was natural for Einstein, and the handful of others prepared to take the light quantum heuristic seriously, to associate the wave picture with the Rayleigh-Jeans regime, and the particle picture to the Wien. They supposed that the wavelike behaviour took place when light quanta were collected together in groups, in the same elementary cells of the harmonic oscillator (this at about the time of the Bohr-Sommerfeld quantisation conditions), or perhaps were guided by waves, explaining also their mutual dependences. For this to happen, light quanta had to be exactly the same, the Planck notion of indistinguishability.[42] But that would seem to be wrong-headed if collections of particles already fail to be mutually independent of each other in the Wien regime, even when particles are rarely if ever found in the same elementary cells.

What, in light of this, are we to think of the complaint of Ehrenfest and Kamerlingh-Onnes? Is it true that Planck's energy elements could not be thought of as Einstein's light quanta, away from the Wien regime? But as just remarked, Einstein's fluctuation probability should then be defined by ratios of symmetric Boltzmann binomials (in the form of Eq.(27)):

$$\frac{(N+V)!}{N!V!} \bigg/ \frac{(N+V_0)!}{N!V_0!}. \tag{36}$$

This is just the expression Einstein would have deduced, with the same exponent $N = E/h\nu$, from the difference in entropy before and after the fluctuation, had he started from Planck's distribution rather than Wien's.[43] Ehrenfest and Kamerlingh-Onnes came within a hairsbreadth of this observation, for they wrote down *precisely* the expression (36), and did so just in order to draw the contrast with (33), as given by Einstein for the Wien regime. Alas, they made no more of it.

## 5. The Bose-Einstein gas

### 5.1 Bose's intervention

There was a good reason why Einstein never took Debye's model seriously, and that is because of its dependence on Maxwell's classical field theory. Debye had thought it a mistake to rely on Lorentz's electrodynamics, as unknown molecular forces appeared to be in play, and for this reason he eschewed use of Planck's resonator equation. He did without it by replacing resonators by modes of cavity radiation, using the Jeans number, and evoking his 'elementary quantum hypothesis'. Einstein agreed with Debye that Planck's resonator equation could not be trusted, but he thought the same of Jeans' calculation (and, one might add, of Debye's quantum hypothesis). And there, more or less, Einstein let the matter lie.

The huge jolt to Einstein's thinking, and what prompted his paper 'The Quantum Theory of the Ideal Monatomic Gas' presented on 10th July 1924 to the Prussian Academy of Science in Berlin, was a letter from a complete outsider, a Mr Satyendra Bose, of the University of Dacca, East Bengal, then 30 years old. He asked for Einstein's assistance in publishing a short paper, written in English, six pages in length. Einstein translated it himself and had it published in *Zeitschrift fur Physik*, and within a week had submitted his own paper.

---

[42] Beginning with the attempts to derive Planck's formula by Wolfke (with Einstein's encouragement – Wolfke was a privatdozent at Zurich in Einstein's Zurich years), and later by Ioffe, Bothe, Ishiwara, and de Broglie. See Darrigol (1991 p.257-60). These efforts were largely driven by a power-series expansion of the Planck distribution in powers of $\exp(-h\nu/kT)$, whereupon it takes the form of a sum of distributions similar to Wien's. I am unconvinced that this line of inquiry contains more than formal comparisons.

[43] In place of Eq.(24), write S as a function of $\rho$. Einstein in 1909 gave what he called a 'similar' treatment to that of 1905 using the Planck distribution, rather than Wien's, but went on to derive the mean square fluctuation of the energy already remarked on; a rather different quantity.

Bose's new idea was mathematically simple, and on a superficial evaluation it was conceptually straightforward: it was to quantise a system by applying the same method of dividing up the one-particle phase space of a system into elementary cells of volume $h^3$, as had been successfully applied to the simple-harmonic oscillator and later by Bohr and Sommerfeld to multiply-periodic systems. In effect, it was to let $\tau = h^3$ in (4) – it could not have been simpler. But on another level it was completely transformative: Bose had calculated the Jeans number as the number of elementary cells in a range of momentum and spatial volume of a one-particle phase space, when the system in question was the light quantum; *for the first time the light quantum was conceived of as a particle with its own state-space*. From this point on we shall use the term 'photon', as it was soon to be called.

It was immediately obvious to Einstein how to apply the same calculation, and obtain the analogue of the Jeans number, for free non-relativistic particles with mass. He fully acknowledged his debt to Bose in the introduction to his own paper, submitted the following week, with the words:

> The path to be taken below, following Bose, is to be described thus: The phase space of an elementary structure (here of a monatomic molecule) is divided, with reference to a given (three dimensional) volume into "cells" of extension $h^3$. If many elementary structures are present, then their (microscopic) distribution as regards thermodynamics is characterised by the ways and means by which the elementary entities are distributed across these cells. The "probability" of a macroscopically defined state (in Plank's sense) is equal to the number of different microscopic states by which the macroscopic state can be thought to be realised. The entropy of the macroscopic state, and therefore the statistical and thermodynamical behaviour of the system, is then determined by Boltzmann's principle. (p.276).

The latter part of Bose's method, as recounted by Einstein, was reasonably familiar (although it involved a certain inversion of the roles of particles and states); it was the first part that was revolutionary. Indeed, in calculating the Jeans number, or what played the role of the Jeans number in the derivation of the black-body formula, in terms of the number of elementary cells in the one-particle state space of the photon, Bose had shown how the one-particle states of the photon could be set up in 1:1 correspondence with the normal modes of the Jeans cube. It was the one-particle state space of the photon, or isomorphic to it. That is, its state space, for finite volume and fixed frequency range, was (isomorphic to) a *finite dimensional vector space*. It is a *vector* space because modes of the radiation field may be superposed.

Events were too fast-moving to know if either Bose, or Einstein, could properly assimilate this idea, or even grasp it; but Schrödinger, by the spring of 1926, surely did, as did Dirac by August that year. But Einstein may have glimpsed it already in the summer of 1924, filtered through de Broglie's ideas, whose PhD thesis had been sent to him by his friend Paul Langevin in Paris. De Broglie had shown how a wave could be associated with a material particle; we shall say more of his influence shortly.

There is the other important aspect of Bose's derivation of the black-body formula: at no point did it so much as hint at the fact that photons, although non-interacting, could not possibly be mutually independent. As in the Solvay model, the Boltzmann binomial did not appear at all, and instead there was only a permutability. Yet the objects in question were (indistinguishable) light quanta, not (distinguishable) Planckian resonators. Nor did Einstein remark on it, neither in his own submission some ten days later, nor in correspondence with Ehrenfest at about the same time; and this despite the fact that what had obstructed the interpretation of thermal radiation in terms of light quanta for almost two decades was the failure of mutual independence in the Rayleigh-Jeans regime. Bose had somehow broken through this difficulty.

He did so through an inversion of the role of particles and states. In place of state distributions, his macrostates specified the number of (distinguishable) elementary cells assigned $k$ (indistinguishable) particles. His microstates, accordingly, did not further specify which particles are assigned to which cell, but rather, for each $k$, which cells are assigned $k$ particles. Historians have widely reported this move as a kind of *mistake*, and Bose's discovery the product of an accidental or serendipitous confusion.[45] It is true that Bose did not comment on the matter. But it is particularly natural following Debye's derivation of Planck's formula (which we know Bose studied), that had replaced Planck's resonators by field modes: if those modes are one-particle states of the photon, then modes of the electromagnetic field, the objects in Debye's treatment that are assigned energy quanta, are turned into one-particle states, that are assigned photons. We have noted the duality operating at the level of the arithmetical identities as well. Einstein himself did not remark on this aspect of Bose's treatment in his summary, just quoted, attributing the concept of macrostate in play to Planck. 'Mistake' is the wrong word for it. But there is no doubt that it concealed the statistical dependence of light quanta, just as effectively as had the Solvay model. It may have seemed that light quanta could be treated as statistically independent after all, even in the Rayleigh-Jeans regime, especially for those convinced by Einstein in 1905 (among them, we must assume, Einstein) that they were mutually independent in the Wien regime.

There is still more to the intrigue. Consider an elementary region $V_s \times [p, p + \Delta p]_s$, indexed by s. Bose's 'inversion' consisted in replacing occupation numbers $n_k^s$ (particle numbers assigned state k) by what are called 'occupancy numbers', integers $p_k^s$ specifying the number of one-partcle states assigned k particles. Macrostates are now complete sets of occupancy numbers $\boldsymbol{p}_s = \{p_0^s, p_1^s, \ldots, \}$. A microstate for Bose specified, of each cell, the number of photons. There are exactly

$$\widetilde{W}_s = \frac{z_s!}{p_0^s! \, p_1^s! \, \ldots} \tag{37}$$

distinct microstates for each macrostate, in Bose's sense of these terms (compare Eq.(17); distinguishable cells are replacing distinguishable particles). He did not mention that this is the correct count of microstates only if interchange of photons does not lead to a new microstate, and that it connects with probabilities only if these microstates are equiprobable.

Eq.(37) is the number of microstates for the macrostate $\boldsymbol{p}_s$ of a single elementary region s. The total number $\widetilde{W}$ for all elementary regions is the product of the numbers (37) for all regions $s$. (Fairly obviously, there is no further multiplicity, corresponding to how the cells are divided up into groups of $z_s$ cells for each elementary region *s*.) The result is:

---

[45] A charge led by Delbruck (1980), followed by Pais (1982) and Darrigol (1991). According to Mehra and Rechenberg (1982 p.567), the change was 'slight' and 'consistent with the spirit of probability theory'. Bach (1990) detects similar shifts in Boltzmann's early combinatorics. Bose was later to say that he was unaware of doing anything out of the ordinary (as recorded by Merha and Rechenberg 1982).

$$\widetilde{W} = \prod_s \widetilde{W}_s = \prod_s \frac{z_s!}{p_0^s! p_1^s! \ldots} \qquad (38)$$

where:

$$\sum_k p_k^s = z_s, \quad \sum_s \sum_k k\varepsilon_s p_k^s = E. \qquad (39)$$

The variational problem proposed by Bose was to maximise (38) on variation of the $p_k^s$ 's subject to the constraints (39). This introduces two Lagrange undetermined multipliers in the way we know and love, the first the normalisation, the second the temperature, with the result:

$$\bar{p}_k^s = A_s e^{-\beta k \varepsilon_s} \qquad (40)$$

with $\beta$ identified as $1/kT$. From Bose's calculation of the number of elementary cells $z_s$ (equal to the Jeans number) the Planck distribution follows immediately.

I have said this follows Debye's method, replacing 'normal modes' by 'elementary cells'; but then why not write down the quantities

$$W_s = \frac{(z_s + p_s)!}{z_s! p_s!} \qquad (41)$$

rather than the permutabilties in (37)? The answer is that Debye had in effect taken a short-cut. On maximising (38), the volume of the resultant equilibrium distribution:

$$\overline{W}_s = \frac{z_s!}{\bar{p}_0^s! \bar{p}_1^s! \ldots}$$

with the equilibrium energy

$$\sum_k k\varepsilon_s \bar{p}_k^s = \bar{E}_s$$

occupies almost the entire available phase space volume. The latter, the total number of microstates in Bose's sense, is:

$$\sum_{p; \sum_k p_k^s = z_s, \ \sum_k k p_k^s = p_s} \frac{z_s!}{p_0^s! p_1^s! \ldots} = \frac{(z_s + p_s - 1)!}{(z_s - 1)! p_s!} \qquad (42)$$

where $p_s = \bar{E}_s / \varepsilon_s$. It is the same arithmetical identity of (18) with (19) written down (without proof) by Boltzmann more than four decades earlier. If we remove reference to s, it is even the same notation. [46] The equilibrium state dominates the sum, so Debye's use of (42) (or, equivalently, (41)) for $W_s$ in the expression to be maximized was bound to yield the same equilibrium entropy.

It is the final piece of the jig-saw. It only remains to step back to see the big picture.

---

[46] We know Bose read Boltzmann (1877) and Debye (1910b). Mehra and Rechenberg agree that he essentially followed Debye (1982 p.565). Bergia's objection (1987 p.244) that if so, he should have used Eq.(42), has just been cleared up. Bose also sent his second paper to Einstein, a few days after his first, in which Eq.(28) figured prominently, with explicit reference to Debye.

## 5.2 Quantum Theory of the Ideal Monatomic Gas

It took two steps back. The first was to understand that Bose's ideas applied just as much to non-relativistic particles as to light quanta, and to particles with a conserved mass (so to particles that clearly changed their states over time). The second was to see that the new statistics built in the statistical dependencies evident for light quanta in the Rayleigh-Jeans regime. Both steps were made by Einstein, the first almost immediately (in his first paper on gas theory, submitted ten days after receiving Bose's manuscript), the second some six months later.

Bose had reasoned that the elementary region $V_s \times [p, p + \Delta p]_s$, of volume $V_s 4\pi p_s^2 \Delta p_s$, degenerate with respect to the energy, should be divided into elementary cells of size $h^3$. The number of cells is then

$$z_s = \frac{4\pi V_s}{h^3} p_s^2 \Delta p_s. \tag{43}$$

In the case of light quanta, as was by 1924 well-known from Compton's treatment of X-ray scattering (but that had figured in Einstein's writings much earlier):

$$\varepsilon_s^2 = p_s^2 c^2 = (h\nu_s)^2. \tag{44}$$

Substituting for $p_s^2 \Delta p_s$ in (43), the result is the Jeans number, Eq.(26), save for a factor of two. In his published paper, this factor was justified by appeal to the two states of polarisation of light.[47] The rest of the derivation was as just outlined, Eqs.(37)-(40).

Einstein applied exactly the same procedure to non-relativistic material particles, making the two obvious needed changes. The first is the replacement of (44) by the non-relativistic energy:

$$\varepsilon_s = \frac{p_s^2}{2m}.$$

Substituting in (43) as before, but expressed in terms of an energy width $\Delta\varepsilon_s$ rather than frequency $\Delta\nu_s$, Einstein obtained:

$$z_s = \frac{2\pi}{h^3} V_s \sqrt{\varepsilon_s} \Delta\varepsilon_s.$$

The second is that mass should be conserved, or equivalently (in non-relativistic physics) particle number should be conserved. There is therefore a new constraint in addition to (39):

$$\sum_s \sum_k k p_k^s = N.$$

The calculation of the equilibrium state then proceeded as with Bose's derivation. The rest, as they say, is history: a new equilibrium state for matter, a new equation of state, and a new phase of matter, the Bose-Einstein condensate.

---

[47] Pais (1986 p.283)) recounts that Bose's English manuscript (subsequently lost) justified the factor two on the hypothesis that light quanta have unit angular momentum that could take only two orthogonal orientations (a version of events supported by Einstein's letter to Bose, 2 July 1924, in Buchwald et al (2015 p.266)). Alas, its translator provided a different rational, referring to the classical concept of polarisation instead.

The introduction of mass was important to a wealth of new physics, but it matters to our story too, because unlike light quanta, it was not in doubt that non-relativistic particles with mass *persisted through change* – that they possessed, at least in a general sense, trajectories. They also possessed a clear state-independent property, preserved in time, namely mass. Plausibly then light quanta did too. It is telling that the term 'photon' was introduced on the assumption that it possessed a state-independent property, namely spin – and, as a matter of record, under the assumption that photon number was conserved.[48] 'Mass zero' was, in due course, to become a state-independent property.[49]

The second step back was to connect to Debye's formulae, and it was made in Einstein's second paper on gas theory, submitted in December 1924. He called Bose's method 'Method (a)' writing down without further comment the volume of an elementary region with $z_s$ cells as:

$$\widetilde{W}_s = \frac{(z_s + n_s - 1)!}{(z_s - 1)! \, n_s!}.$$

and for the total number of microstates the product of these Boltzmann binomials over *s*, Eq.(15), yielding the entropy

$$\tilde{S} = \sum_s \{(n_s + z_s) \ln(n_s + z_s) - n_s \ln n_s - z_s \ln z_s\}. \tag{45}$$

Einstein immediately concluded:

> It is easily recognised that by this calculation approach the distribution of molecules over the cells is not treated as statistically independent. This is because the cases, here called 'complexions', would not be regarded of equal probability according to the hypothesis of an independent distribution of the individual molecules among the cells. For really statistically independent molecules, the counting of these 'complexions' of different probabilities would not yield the entropy correctly.

Indeed, away from the limit $z_s \gg n_s$, probabilities are affected even at the level of individual molecules. Thus for $z_s = n_s = 2$, the probability of a single molecule in each cell is one-third (as for the only two other microstates, where both are in the same one-particle cell); whereas if the interchange of particles defines a new microstate, it is one-half, assuming microstates are equiprobable. The latter point is the same as Natanson's critique.

In contrast, 'Method (b)' is defined 'according to the hypothesis of the statistical independence of the molecules', with volume measure (12):

$$W_s = z_s^{n_s}$$

and for the total number of microstates for all the elementary regions Eq.(14):

---

[48] Both were required by G. N. Lewis in 1926, when he coined the term 'photon'. It was moreover an property invariant under Lorentz transformations ('all photons are alike in one property which has the dimensions of action or of angular momentum, and is invariant to a relativity transformation' (Lewis 1926 p.874)).

[49] As prefigured in the abstract of Einstein's third and final paper on gas theory, 'This theory seems legitimate when one starts from the conviction that a light quantum (disregarding its polarisation property) differs from a monatomic molecule essentially only in that the quantum's mass if vanishingly small' (Buchwald et al, p.418).

$$W = \frac{N!}{n_1! \dots n_s! \dots} \prod_S W_s$$

including the overall permutability for exchanges of particles between different elementary regions. The entropy is:

$$S = kN \ln N + k \sum_S \{(n_s \ln z_s - n_s \ln n_s)\}. \qquad (46)$$

Einstein had almost the full picture; only absent was the limiting equality Eq.(11), for $z_s \gg n_s$ -- given which it would have been obvious that

$$\widetilde{W} \approx \frac{W}{N!}, \qquad (47)$$

the equation written down by Tetrode in 1914, in a notation that made clear his debt to Gibbs, and that drew published criticism from Lorentz.

Einstein did not just omit this observation; he missed it. For commenting on the entropy (46), obtained using Method (b), he said of the leading term $N \ln N$:

> When comparing the entropies of different macroscopic states of the same gas, this term plays the role of an inconsequential constant that we can leave out. We *must* leave it out if -- as is customary in thermodynamics -- we want to achieve that the entropy be proportional to the number of molecules of a given inner state of the gas. …One usually tends to justify this omission of the factor N! in W for gases by regarding complexions arising from the mere exchange of molecules of the same kind as not different and, therefore, as being counted only *once*. (p.375).

But that is precisely the assumption of the new statistics, Bose's method, as would have been obvious from (47). Had Einstein seen that division by $N!$ follows as the dilute limit of the new statistics, using Bose's method, he would have said so.[50] That Method (b), corrected by division by $N!$, is the dilute limit of Method (a), likewise went unmentioned.

Were the connection clear, Einstein would also have noted it in light of Nernst's theorem, the so-called third law of thermodynamics: the requirement that the entropy of a gas goes to zero with the absolute temperature. If a simple subtraction of the term $kN \ln N$ is made from (46), Einstein noted, the resulting entropy function no longer satisfies the third law. At absolute zero, all particles will be in the first quantum state:

$$n_s = 0 \text{ for } s \neq 1$$
$$n_1 = N \qquad (48)$$
$$z_1 = 1.$$

But the corrected entropy function is then $-kN \ln N$, rather than zero. As Einstein observed, in the new gas theory, using Method (b), the choice between extensivity and the Nernst theorem is exclusive: you cannot have both. But you have both for the entropy $\widetilde{S}$. (How so, in light of Eq.(11), when the latter yields the 'corrected' entropy function for statistically-independent particles? – because Eq.(48) is about as far from the limit $z^s \gg n^s$ as you can get. Einstein was silent on this point as well.) Perhaps surprisingly, since he had never before allowed the 'inconsequential constant' $kN \ln N$ to cause him any trouble,[51] Einstein concluded that the statistical independence of molecules must be given up for the sake of extensivity:

> For these reasons I believe that calculational approach (a) must be given preference, although the preference for this calculational approach over other approaches cannot be proven a priori. This result, for its part, constitutes a support for the notion of a deep essential relationship between radiation and gas, in that the same statistical method that leads to Planck's formula establishes agreement between gas theory and the Nernst theorem in its application to ideal gases. (p. 376).

---

[50] The phrase 'one usually tends to justify' suggests the inference is unsound; this, I suggest, is a faint echo of Lorentz's complaint about Tetrode.

His volte face on the extensivity puzzle may have been opportunistic. Einstein had, after all, already gone into print with his first paper on the gas theory (it was published 20[th] September 1924), whereupon arguments subsequently found to be in support of it were naturally welcome. But Einstein had in his hands a new *quantum* theory of the ideal gas, whether on the hypothesis of Method (a) or Method (b). Nernst's theorem required an absolute value of the entropy for the ideal gas, and he (rightly) thought it a quantum phenomenon; that surely counselled against ad hoc subtractions as a way of solving the $N!$ puzzle, subtractions that might be tolerated in a theory, like classical statistical mechanics, known to be unsafe, but not in a new quantum theory of gases. The alternative was Method (a).

## 5.3 Waves and Particles

Einstein could not, however, quite bring himself to admit that light behaves as if it is a gas of non-interacting light quanta across the entire frequency spectrum (and not just in the Wien regime). He preferred, somewhat contrary to his conclusion as just stated, to import into the new gas theory the many features that had troubled him for so long concerning black-body radiation. Following what was in essence Natanson's critique, he went on to say:

> Consequently, the formula indirectly expresses a certain hypothesis about an initially completely puzzling mutual influence of the molecules that determines just the same statistical probability of the cases that are defined here as 'complexions' (p.374).

He drew attention to the other fluctuation formula already referred to, now written for the mean square of the difference $\Delta_s = n_s - \bar{n}_s$. It can be directly derived from the entropy (45), so it applied just as much in the new gas theory as in black-body radiation. The result is:

$$\overline{\Delta_s^2} = \bar{n}_s + \frac{\bar{n}_s^2}{z_s} \ .$$

Einstein renewed his interpretation of the two terms he had given more than fifteen years before: the first as follows for independent molecules, and the second as follows for classical radiation, adding: 'One can interpret it likewise in the case of a gas by suitably assigning to the gas a radiation process and calculating its interference fluctuations' and 'I shall delve deeper into this interpretation because I believe it involves more than a mere analogy' (p. 377).

---

[51] See Darrigol (2018 p.30) for Einstein's few comments on the extensivity puzzle.

Enter de Broglie. Einstein continued 'how a material particle or a system of material particles, can be assigned a (scalar) field of waves has been demonstrated by Mr de Broglie in a very noteworthy paper', adding in a footnote that 'there is a very remarkable geometrical interpretation of the Bohr-Sommerfeld quantum rule in this dissertation as well'. He went on to sketch de Broglie's relation between phase and group velocities for a particle of mass $m$, with an associated frequency $mc^2 = h\nu$. Einstein, in short, pushed for an explanation for the failure of statistical independence in terms of a new kind of kinematics.

Insofar as there was something right about de Broglie's ideas, there was something right about this instinct of Einstein's. And of course there *was* something right about de Broglie's ideas, because they *did* lead Schrödinger to the wave mechanics. Schrödinger had a long-standing interest in gas theory, and in February 1925 he wrote to Einstein on the failure of statistical independence that he had detected in Einstein's first paper on gas theory. Einstein in reply directed him to his (just published) second paper, and thereby to de Broglie's thesis. Schrödinger's two papers that followed, both on quantum gas theory, immediately proceeded the first of his historic papers on wave mechanics.

This is fast-moving history and it has received plenty of attention by historians,[52] but again it has a special meaning from the point of view of the concept of indistinguishability. The first of Schrödinger's two papers on quantum statistics continued a debate with Planck on whether quantisation should be applied to a gas as a whole (states on $6N$-dimension phase space), or as a discretisation of one-particle phase space, and on various definitions of the entropy in relation to the extensivity of the entropy, which Planck had previously justified on the basis of Gibbs' concept of generic phase. Including as it did commentary on the new Bose-Einstein definition of the entropy, this was the closest that anyone came to making the connection between quantum statistics and Gibbs' concept of indistinguishability, prior to Epstein in 1936, but still Schrödinger did not mention Gibbs by name, did not reference the *Principles*, did not mention the generic phase.[53]

The second of Schrödinger's papers offered an alternative to the Bose-Einstein theory. With hindsight it was a retrograde step: it did not so much explain the correlations, as abolish them, along with the particles. 'The real, statistically independent entities' were now modes of a matter field, but it was the state of the entire field that was subject to a discretisation of the energy, not the individual modes. But for this difference, it was a near equivalent to Debye's theory of the radiation field, played out in terms of a matter field. Schrödinger did not present his new quantized matter field as *equivalent* to a many-particle theory, as hinted at by Einstein, and as established by Dirac in 1927; he sought a rival theory.[54] His aim was to get rid of the many-particle viewpoint altogether, and with it the violation of statistical independence

Schrödinger's 'On Einstein's gas theory' was submitted on December 15th 1925. Was it the basis for his first paper on wave mechanics, received March 23 1926? Or was it that from mid-December he was finally free to consider other aspects of de Broglie's thesis? – most notably the one Einstein had called not just noteworthy, but *remarkable*: the derivation of the Bohr-Sommerfeld quantisation conditions on the basis of an allowed set of stationary waves. Recall the title of Schrödinger's great paper of March 23, 'Quantisation as an Eigenvalue Problem'; it made no connection with quantum statistics.

---

[52] Notable among them Hanle (1977), Mehra and Rechenberg (1987 p.361-65, 367-402), and Darrigol (1986).
[53] Planck responded with an address to the same meeting of the Prussian Academy of Sciences, on 23 July 1925, at which Schrödinger's paper was presented. See Mehra and Rechenberg (1987 9.393-97), Darrigol (1991, p.295-98) for further discussion.
[54] Indeed, as Schrödinger pointed out, his theory did not contain the Bose-Einstein condensate.

Where, when the dust had finally settled, did it all end? Was Einstein right that indistinguishability shows a mysterious influence, a correlation, and is it explained in the new wave mechanics, or in quantum field theory, and if so in terms of what? There are many historians and philosophers of physics who think the answer is yes to both – and that the explanation is *quantum entanglement*. According to them, Einstein was groping his way to a mysterious kind of influence that he was eventually to distil in the EPR paradox and in the idea of non-separability, that we *know* is due to entanglement.[55] There is a prima facie case to conclude that the failure of statistical independence is indeed a quantum phenomenon and unlike anything encountered in classical physics. And in that case, the parallel with Gibbs notion of indistinguishability as defined by the generic phase cannot possibly be as good as it seems.

Against this, however, there is a growing body of evidence to show that the particular kind of entanglement required by symmetrisation is of an essentially trivial kind. Not only is it consistent with the existence of well-defined particle trajectories (meaning, orbits of one-particle states under the unitary evolution, as we shall see in a moment), but it is also insufficient to violate any Bell inequality.[56] It is, rather, just what is required to ensure factor-position plays no dynamical role in the theory, or show up on any prediction of the theory, provided the Hamiltonian is a symmetric function of particles coordinates (and if it is not: then it does not preserve the symmetrisation of the state). In illustration, the partial trace of a symmetrised state is the same regardless of the factors traced over, a circumstance that has led some otherwise distinguished commenters to conclude that every indistinguishable particle is in exactly the same state.

Symmetrisation says only that particles cannot be assigned states on the basis of factor position; they may yet be identified with one-particle states, as Gibbs' generic phase shows in the classical case.

## 6. Epilogue

### 6.1 The Gibbs Paradox Revisited

Why is it that on the diffusion of two samples of the same gas at the same temperature and pressure into one another, the resulting entropy is not the same as the sum for the two taken separately – even when the gases are made of non-interacting particles? Why is there an entropy increase when only one chamber is occupied, but not when both are occupied, and how is it they then cancel? The answer is because when both are occupied, what would otherwise be new many-particle states, made available by the removal of the partition, *were already available*; they have already been taken into account. The reason for the lack of statistical independence of the two gases, after mixing, is that whilst there are plenty of new *motions* available to the particles – they may now pass from region *A* to region *B*, and vice versa– but there are no (or almost no) new *N*-particle *states*: there are only new ways of leaving and arriving at the very same states. (The qualification 'almost no' is needed because with the partition removed, different numbers $N_A{'}$, $N_B{'}$, summing to $N$, may be found in $A$ and $B$ respectively, so there undoubtedly are new states in $\tilde{\Gamma}$ that are collectively accessible; but they amount only to fluctuations in number density, contributing a negligible increase to the entropy, zero in the Sterling approximation.)

---

[55] For a persuasive case of this kind, see Howard (1990), endorsed also by Norton (2006).
[56] See Caulton (2020), Ghirardi et al (2002), (2004).

In illustration, in Fig.2, the initial and final states of the two particles are numerically identical as points in $\tilde{\Gamma}$; the two particles return to their initial state. Although the state of each particle, before and after, has changed (from $(p,q)$ to $(p',q')$ and from $(p',q')$ to $(p,q)$), the instantaneous state of the two particles taken together, before and after, is the same. I say this as an ontological claim, not an epistemic one: the microstates are identically the same.

Was this implication of Gibbs' concept of generic phase, taken realistically, glimpsed by its critics? And if it was, was it deemed a contradiction in terms? I know of no discussion of the matter. It may have seemed like one (how can the state of each of two things be changed, yet the state of the two be the same?), yet there is no *logical* contradiction here. But it is much easier to accept the epistemic notion: that at *some level of coarse-graining*, the results of particle interchanges are the same. This may have been enough for Debye and Planck, but not for realists like Ehrenfest, Einstein, Lorentz, and Schrödinger.[57] It would not explain the statistical dependencies, unless the dynamics, too, is blind to finer scales.[58]

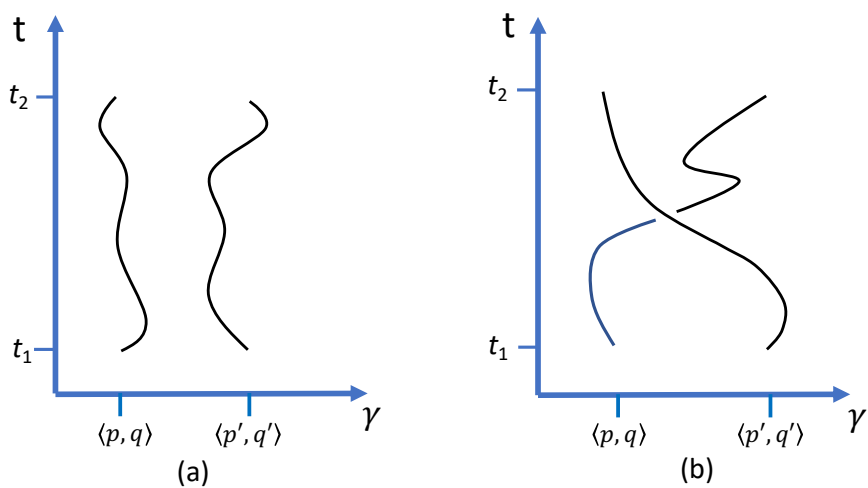

Figure 2: The final state in (a) and (b) is the same.

## 6.2 Dirac's Argument for Symmetrisation

The radical alternative to all this is that an interchange of quantum particles over time is not a real physical process at all. It rapidly became orthodoxy, just as 'all this' might finally have emerged from the shadows. In Paul Dirac's first foray into wave mechanics, 'On the theory of quantum mechanics', submitted to the *Proceedings of the Royal Society of London* in August 1926, such transitions seemed to be denied.

---

[57] Rosenfeld (1959) puts Gibbs in the idealist camp as well.
[58] See Saunders (2013, 2018) for more in this vein.

The paper, even by the standards of the young Dirac, was a tour de force. It showed that the invariance of the state under particle interchange could be defined in two, mutually exclusive ways: either by symmetrisation or antisymmetrisation of the vector-state (or wave-function). In the former case it is unchanged, whereas for the latter it alternates in sign, depending on whether the permutation involves an even or odd number of particle interchanges. Both leave the *state* unchanged (for the state is the vector-state up to phase). The former leads to Bose-Einstein statistics (here Dirac cited Bose's paper, and the two papers by Einstein we have considered), whereas the latter leads to what was soon to be called Fermi-Dirac statistics (here Dirac cited Pauli). It showed that particles with antisymmetrised wave-functions satisfy Pauli's exclusion principle, and that the Hamiltonian, if it is to define a unitary evolution on the space of symmetrised states, must be a symmetric function of the particle positions and momenta.

Novel though all this must have seemed at the time, we have prepared enough of the ground to see its classical counterparts using the generic phase. The distinction between symmetric and antisymmetric states is that between generic phases where particles can occupy the same one-particle states or points, and where they cannot – whether on taking the quotient space under permutations, coincidences of coordinates (in the case of Figure 1, the one-dimensional boundary $x_1 = x_2$) should be excised. The difference matters to the topology of the state-space $\tilde{\Gamma}$, but makes no difference to the volume measure. In quantum mechanics it is the other way round. All three state-spaces, the space of states of distinguishable particles, the space of symmetrised wave-functions, and the space of antisymmetrised wave-functions, have the same topology (they are all subspaces of Hilbert space, closed in the norm topology), but as is expected for a finite dimensional state space, each has a different 'volume' (dimension). Thus let the one-particle Hilbert space $\hbar$ have dimension Z. Then the dimension of the N-fold symmetrisation of $\hbar$, for symmetric wave-functions, is

$$\frac{(Z+N-1)!}{(Z-1)!\,N!}.$$

It is our last look at the Boltzmann binomial. The dimension for antisymmetric wave-functions is instead:

$$\frac{Z!}{(Z-N)!\,N!}$$

as can also be obtained by a simple adaptation of the symbol-sequence argument. The two agree in the classical limit $Z \gg N$, and with Gibbs' generic phase.

But Dirac did not consider the classical limit, and he did not make the connection with Gibbs' ideas. He did not so much ignore the links with classical statistical mechanics as to deny them: the new mechanics was to be developed on the principle that only measurable quantities were to have a mathematical meaning. He wrote:

> In Heisenberg's matrix mechanics it is assumed that the elements of the matrices that represent the dynamical variables determine the frequencies and intensities of the components of radiation emitted. The theory thus enables one to calculate just those quantities that are of physical importance, and gives no information about quantities such as orbital frequencies that one can never hope to measure experimentally. We should expect this very satisfactory characteristic to persist in all future developments of the theory.

Frequency is determined by the energy difference of the states ('orbits') involved in the transition: thus $m \to n$ is represented by a matrix element, its modulus square is the intensity, and $v = (E_m - E_n)/h$ is the associated frequency. To enquire into the orbital velocities of the electrons themselves, about the nucleus, as fruitlessly pursued by Dirac prior to Heisenberg's revolutionary paper of 1925, had obviously been a mistake, and he had taken the lesson to heart:

> Consider now a system that contains two or more similar particles, say, for definiteness, an atom with two electrons. Denote by $(mn)$ that state of the atom in which one electron is in an orbit labelled $m$, and the other in the orbit $n$. The question arises whether the two states $(mn)$ and $(nm)$, which are physically indistinguishable as they differ only by the interchange of the two electrons, are to be counted as two different states or as only one state, i.e. do they give rise to two rows and columns in the matrices or to only one? If the first alternative is right, then the theory would enable one to calculate the intensities due to the two transitions $(mn) \to (m'n')$ and $(mn) \to (n'm')$ separately, as the amplitude corresponding to either would be given by a definite element in the matrix representing the total polarisation. The two transitions are, however, physically indistinguishable, and only the sum of the intensities for the two together could be determined experimentally. Hence, in order to keep the essential characteristic of the theory that it shall enable one to calculate only observable quantities, one must adopt the second alternative that $(mn)$ and $(nm)$ count as only one state.

The transition $(mn) \to (m'n')$ corresponds to Fig.3a, the transition $(mn) \to (n'm')$ to Fig.3b, where the horizontal axis parameterises the different orbitals, or stationary states, of the electrons (and the vertical axis is as before the time). The two end states are the same 'because the two transitions are physically indistinguishable', so we should not be able to calculate their separate contributions – they should not, indeed, occur as distinct elements in the theory at all, on Dirac's operationalist philosophy. Even realists like Schrödinger agreed: twenty years later, writing on Gibbs' paradox in his book *Statistical Thermodynamics*, the paradox is solved in quantum theory 'because exchange of like particles is not a real event – if it were, we should have to take account of it statistically' (p.61).

Contrast the classical case, and with it Gibbs' generic phase: classically there invariably *are* facts of the matter, as determined by the dynamics, as to whether or not the process of Fig.2a took place, or Fig.2b; and this is so whether or not they can be used to calculate any measurable quantity. The continuous motions of classical particles around each other leading to the interchange of their positions and momenta is a real physical process. We are back to particle trajectories; it is this argument of Dirac's, just quoted, that cemented the doctrine: quantum indistinguishable particles cannot have trajectories.

## *6.3 Quantum Trajectories*

Except that they can. The clinching argument for sameness of the classical and quantum concepts is that particle trajectories can perfectly well be defined in quantum mechanics as they can classically, even for symmetrised states. That may not be possible in the processes considered by Dirac, but it clearly is for others, at least for non-interacting particles.

Consider a symmetric vector state of two identical particles of the form:

$$|\Psi\rangle = \frac{1}{\sqrt{2}}(|\varphi_m\rangle \otimes |\varphi_n\rangle + |\varphi_n\rangle \otimes |\varphi_m\rangle). \qquad (49)$$

We may perfectly well interpret this in terms of the state $|\varphi_m\rangle$ of one quantum particle, and the state $|\varphi_n\rangle$ of another, without having to say which particle has which state, in exactly the same way as given the unordered pair $\{\langle q_m, p_m\rangle, \langle q_n, p_n\rangle\} \in \tilde{\Gamma}$ we interpret $\langle q_m, p_m\rangle$ as the position and momentum of one classical particle, and $\langle q_n, p_n\rangle$ as the position of another.[59] Missing in both cases is any prior standard of which particles we are talking about, as was provided by factor-position in the tensor product of Hilbert spaces $\hbar$ for unsymmetrised states, and respectively in the Cartesian product of phase spaces $\gamma$ for the specific phase. If the unitary evolution of $|\Psi\rangle$ does not lead to *superpositions* of such states – if it is of the form $\widehat{U}_t \otimes \widehat{U}_t$ – then there is a one-particle state $|\varphi_m\rangle$ evolving under $\widehat{U}_t$, and another one-particle state $|\varphi_n\rangle$ evolving under $\widehat{U}_t$; and the two evolutions may be quite different, as beginning with quite different states.

This recipe extends without any difficulty to $N$-particle symmetrised states, including states in which many particles are in the same state. It fails only when *superpositions* of symmetrised states are considered, that is given *genuine* entanglement (as opposed to the trivial kind involved in symmetrisation of the state). But the inability to attribute definite properties to quantum particles when genuinely entangled is a general feature of quantum mechanics, that applies as much to superpositions of product states of distinguishable particles as to superpositions of symmetrised states of indistinguishable particles. It is independent of the concept of indistinguishability. It is not *indistinguishability* that leads to the absence of trajectories, it is the unrestricted validity of the superposition principle.

Particle interchanges, actively interpreted, can be defined in the quantum case just as in the classical: Fig.2a and 2b may well arise as different quantum processes, because they may arise under different Hamiltonians, just as in classical theory. Thus, consider the one-particle unitary evolution $\widehat{U}_t$ on $\hbar$ with the action:

$$\widehat{U}_t: |\varphi_m\rangle \to |\varphi_{m'}\rangle; \; \widehat{U}_t: |\varphi_n\rangle \to |\varphi_{n'}\rangle$$

corresponding to Dirac's transition $(mn) \to (m'n')$. For the transition $(mn) \to (n'm')$ we have the unitary $\widehat{U}_t^\pi$ acting as

$$\widehat{U}_t^\pi |\varphi_m\rangle = |\varphi_{n'}\rangle; \; \widehat{U}_t^\pi |\varphi_n\rangle = |\varphi_{m'}\rangle.$$

In terms of unsymmetrised states, Fig.3a and 3b correspond to the two evolutions:

$$\widehat{U}_t \otimes \widehat{U}_t: |\varphi_m\rangle \otimes |\varphi_n\rangle \to |\varphi_{m'}\rangle \otimes |\varphi_{n'}\rangle$$

$$\widehat{U}_t^\pi \otimes \widehat{U}_t^\pi |\varphi_m\rangle \otimes |\varphi_n\rangle \to |\varphi_{n'}\rangle \otimes |\varphi_{m'}\rangle.$$

The two are quite different; it is no longer obvious that the final states can be identified. Nevertheless, they *should* be identified – not, as in the classical case, by passing from ordered pairs of states to unordered pairs, but by symmetrisation. This is perfectly compatible with

---

[59] For antisymmetrised states a 'preferred basis' is needed – preferred, that is, by decoherence theory. See Saunders (2013, 2016) for further discussion.

the difference between Fig. 3(a) and 3(b): given the initial 2-particle symmetrised state (49), the dynamics

$$\hat{U}_t \otimes \hat{U}_t : |\Psi\rangle \to |\Psi'\rangle$$

corresponds to Fig.3a, and

$$\hat{U}_t^\pi \otimes \hat{U}_t^\pi : |\Psi\rangle = |\Psi''\rangle$$

to Fig.3b; these are two distinct dynamical processes, but they yield identically the same state $|\Psi'\rangle = |\Psi''\rangle$, as can be seen by inspection.

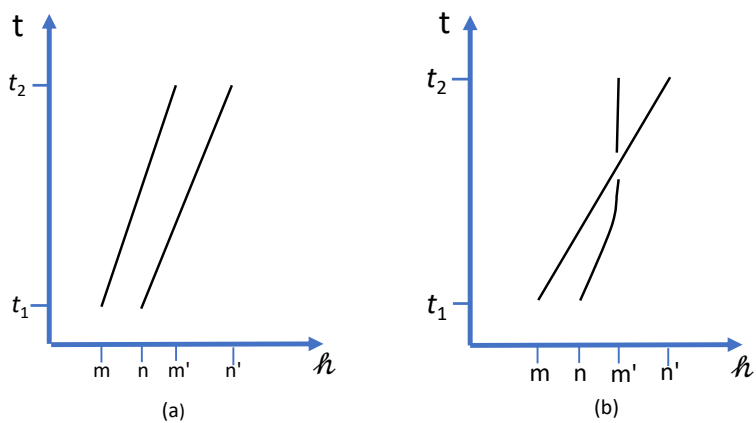

Figure 3: Dirac's two transitions (a) $(mn) \to (m'n')$, (b) $(mn) \to (n'm')$

The argument for particle trajectories can even be run in an approximately classical sense: when the one-particle states in question $|\varphi_m\rangle$, $|\varphi_m\rangle$, are well-localised in configuration space and in momentum space, throughout the time in question – namely, when the states are Gaussians, with small spreads in momentum (this is much easier to arrange when the particles are relatively heavy).[60] That is, quantum indistinguishable particles may well have approximate, quasi-classical trajectories, that can be arranged in one of two ways, differing in which particle ends up in which one-particle state. Dirac's argument for identifying the joint final states in the two cases then fails, for now there *is* something in the formalism to say which transition took place – namely, the trajectories. Yet the two should still be identified, here as in the classical case.

**Acknowledgment**:

I would like to thank the Leverhulme Trust for their generous support in the writing of this article.---

[60] See e.g. Wallace (2012, pp.64-67).